\documentclass[11pt]{article}
\usepackage{amsmath,amssymb,mcite, citesort}
\usepackage{rotating}
\usepackage{caption}
\renewcommand{\d}{\textup{d}}
\newcommand{\e}{\textup{e}}

\begin{document}

\begin{titlepage}

\hfill AEI-2013-130

\hfill DAMTP-2013-12 

\vspace{0.75cm}
\begin{center}

{{\LARGE  \bf 	$E_{8}$ duality and dual gravity}} \\

\vskip 1.5cm {Hadi Godazgar$^{\star}$ 
, Mahdi Godazgar$^{\dagger}$ 
and Malcolm J. Perry$^{\ddagger}$}
\\
{\vskip 0.5cm
$^{\star \dagger}$ Max-Planck-Institut f\"{u}r Gravitationsphysik, \\
Albert-Einstein-Institut,\\
Am M\"{u}hlenberg 1, D-14476 Potsdam, Germany
\vskip 0.5cm
$^{\ddagger}$ DAMTP, Centre for Mathematical Sciences,\\
University of Cambridge,\\
Wilberforce Road, Cambridge, \\ CB3 0WA, UK\\}
{\vskip 0.35cm
$^{\star}$ hadi.godazgar@aei.mpg.de, $^{\dagger}$ mahdi.godazgar@aei.mpg.de, 
$^{\ddagger}$ m.j.perry@damtp.cam.ac.uk}
\end{center}

\vskip 0.35cm

\begin{center}
\today
\end{center}

\noindent

\vskip 1cm

\begin{abstract}
\noindent We construct the non-linear realisation of the $E_{8}$ motion group and compare this with the bosonic sector of eleven dimensional supergravity. The construction naturally leads to the introduction of a new potential field.  We identify this new field with the dual gravity field by considering the reduction of the eleven-dimensional theory to three dimensions.
\end{abstract}

\end{titlepage}

\section{Introduction}

\noindent The toroidal compactification of eleven-dimensional supergravity \cite{CJS} to various dimensions leads to hidden symmetries \cite{cremmerjulia, *so(8), julia, nicolaie9, cremmerjulialupope1}, which have influenced many important developments. Arguably, they have played an integral part in the set of ideas leading to U-dualities and the conjecture of M-theory \cite{mtheoryht, mtheoryw}. Furthermore, they continue to provide insights into a wide-range of problems associated with string/M-theory. However, the role of these symmetries in the full eleven-dimensional theory remains unclear. While these symmetries only appear upon reduction,  early seminal work \cite{dewitnicolai, nicolai} found evidence that these symmetries are not merely artifacts of the reduction. They showed that eleven-dimensional supergravity can be reformulated in a way that makes the local symmetries associated with the global exceptional symmetries $E_7$ and $E_8$, respectively, manifest. Moreover, they were able to assemble some bosonic degrees of freedom into representations of the global symmetry groups. While eleven-dimensional supergravity does not admit the global symmetries, these work hint at structures in eleven dimensions that naturally encompass the duality structure that appears under reduction. Such a framework would provide a more direct understanding of these duality symmetries from an eleven dimensional perspective and possibly shed light on M-theory.

More recent attempts in trying to understand these duality symmetries have centred on generalised geometry \cite{hitchin1, hitchin2, gualtieri, westl1, hullm, PW} and related ideas of exceptional geometry \cite{KNS, hillmann}. These ideas are based on the extension of the tangent space of a geometry to include $p$-form bundles, and in some cases, also an extension of the base space to include dependence on new coordinates that are seen as windings of branes. The extension of the space geometry to include windings associated to the branes leads to the unification of gravity and the fields sourced by the branes in a single description \cite{davidmalcolm, BGP, BGPW}. In this approach, the dynamics of fields along the internal directions are formulated in terms of a generalised metric that is found from membrane duality arguments \cite{DuffLu} or  constructed from the duality coset. Thus rendering the description duality-manifest. 

In the context of string theory, similar considerations have been made with respect to the T-duality group.  In the double field theory \cite{dft1, dft2, dft3, genmetstring} approach to closed string theory, motivated by string field theory, all fields are taken to depend on dual (winding) coordinates as well as spacetime coordinates. This naturally leads to a generalised geometric structure in which the extended diffeomorphism contains both spacetime diffeomorphism and the gauge symmetry of the NS-NS 2-form.  The generalised diffeomorphism algebra closes on the assumption that generalised fields satisfy a differential constraint, known as the section condition, that reduces their dependence to a subset of coordinates.  From a physical point of view, the section condition is the level matching condition in string theory.  While this geometry does not admit some familiar notions of differential geometry, such as the usual concept of a connection, it does possess a structure \cite{siegel1, *siegel2}\cite{
genmetstring} \cite{gendiff1, gendiff2, gendiff3, CSW} that in particular contains analogues of the Ricci tensor and scalar---the equation of motion and Lagrangian of the low-energy effective description of closed string theory.  The generalised geometric descriptions of heterotic \cite{dfthet1, dfthet2} and type II theories \cite{dfttypeii1}\cite{CSW}\cite{dfttypeii2} also exist.

The generalised geometries associated with the M-theory dualities admit similar, but richer structures given the existence of higher rank $p$-forms sourced by various branes \cite{hullm}.  As with double field theory, the generalised geometries in this context also contain notions of a generalised diffeomorphism algebra that unifies spacetime diffeomorphims and gauge symmetries and closes on a section condition \cite{BGGP, CSWex, BCKT}, as well as other structures \cite{CSWex2, Ugeom, CEK, AGMR}.

In \cite{BGPW}, the SL(5), SO(5,5) $E_{6},$ and $E_7$ duality groups were considered and the dynamics of the corresponding internal fields were described by a non-linear realisation \cite{ISS, bo, west2000, locale11} of the respective groups seen as subgroups of $E_{11}.$ The focus of this paper is the non-linear realisation of $E_8.$ This is the duality group of maximally supersymmetric three-dimensional supergravity \cite{MSe8} that appears upon the toroidal reduction of eleven-dimensional supergravity \cite{juliae8}.

As eluded to earlier, a reformulation of eleven-dimensional supergravity with respect to the $E_8$ duality group was first considered by Nicolai in \cite{nicolai} and elaborated on further in \cite{KNS}.  In particular, in \cite{KNS}, the authors provide evidence for a `generalised vielbein' in the $248 \otimes 248$ of $E_8$ and the unification of spacetime and gauge symmetries in the internal directions.  The study of supersymmetry
transformations and the treatment of the 3-form potential as an independent field is central to their argument and what emerges is a structure that can be viewed as belonging to the $E_8$ tensor product representation $36 \otimes 248$. From the perspective of this work, the failure of a generalisation of this structure to a $248 \otimes 248$ object there is due to the absence of dualisation of relevant fields.  Therefore, to understand the significance of duality symmetries in the eleven-dimensional theory, it is necessary to have in mind a `democratic formulation' in which the supergravity fields are supplemented by their duals, i.e.\ the 6-form and the dual gravity field \cite{cremmerjulialupope1}.  This is not so surprising when viewed from the perspective of the reduced theory and the necessity of dualisations for the appearance of symmetries.  Furthermore, dualisation is necessary in the local SU$(8)$ \cite{dewitnicolai} and SO$(16)$ \cite{nicolai} invariant reformulations of eleven-dimensional 
supergravity.

Whereas the dualisation of form-fields is well-understood, the dualisation of the metric field is more intricate. The interpretation of the curvature tensor as a 2-form field strength of the metric field allows for a natural generalisation of dualisation in this context \cite{dualHull1, dualHull2, weste11, dualHull3, HTdualgrav}. In the linearised theory, this leads to actions for gauge fields in exotic representations of the Lorentz group first considered by Curtright \cite{Curtright}. It has been argued \cite{BBH}  that the extension of such an idea to the non-linear theory is not possible in a local and covariant manner, in general. Although, the existence of isometries is one way to circumvent this \cite{dualHull3}.  In such a setting, the dual gravity field is the dual of the graviphoton gauge field. The relevance of a dual gravity formulation in the context of M-theory dualities \cite{OPR, OP}, in particular the $E_{11}$ proposal, has been of much recent interest \cite{BdRKKR, boulangerhohm, dem, 
westdual, 
offshelldual}. In these papers the possibility of introducing dual gravity fields transforming under the gauge symmetries of the matter fields of eleven-dimensional supergravity has been investigated. In particular, in \cite{westdual} the dependence of the dual gravity field on the 3-form gauge field and its dual has been predicted from $E_{11}.$
  
The $E_8$ duality group is particularly interesting from the point of view that the potential of the dual gravity field is expected to appear in the generalised metric for the first time.\footnote{Winding coordinates that can be interpreted as those of a Kaluza-Klein monopole do appear in the $E_7$ algebra, but the potential associated with these coordinates does not appear in the generalised metric.} As has been observed in \cite{CSWex, BCKT}, the presence of dual gravity poses difficulties for the formulation of an $E_8$ generalised geometry.  

The goal of this paper is to construct the non-linear realisation of the $E_8$ group and compare it with what one would expect from the bosonic sector of eleven-dimensional supergravity. In section \ref{genmet}, we begin by constructing the non-linear realisation of the $E_8$ motion group.  The main steps in this construction are as follows.
\begin{itemize}
 \item We ascertain the $E_8$ motion group, which is the semi-direct product of the $E_8$ group with that of its adjoint representation.  The adjoint representation can be thought of as being generated by translations.  This is analogous to the definition of the Poincar\'{e} group as the the semi-direct product of the Lorentz group with that of its vector representation, the elements of which are viewed as translation generators.  The $E_8$ motion group is given in terms of an SL(8) decomposition of the $E_8$ algebra and its adjoint representation.  This is because from an eleven-dimensional perspective, the $E_8$ duality group appears in the reduction to three dimensions on an 8-torus.  Thus, we would like the duality group to act on the eight spatial directions that would be associated with the torus under the reduction.  
 \item We construct the generalised $E_8$ vielbein by conjugating the Maurer-Cartan form of an element of the adjoint representation with an element of the $E_8$ group.  This is equivalent to calculating the Maurer-Cartan form of an element of the motion group and reading off the part that appears as a coefficient of the translation generators.  Given its transformation properties, this object defines a vielbein.
 \item We formulate the $E_8$ invariant dynamics for the eight-dimensional space in a canonical approach.  In such a description, the dynamics is given by a potential and a kinetic term.  The strategy in this construction is to write down all $E_8$ invariant terms constructed from the generalised metric and fix their coefficients by requiring that the expression reduces to what one would expect for the gravitational sector.  Once the coefficients are fixed the full expression with all fields turned on can be computed with the assumption that fields do not depend on the generalised coordinates.
 \end{itemize}

The potential term that is obtained includes an Einstein-Hilbert term, which appears by construction; gauge-invariant field strengths of a 3-form ($C_{abc}$) and a 6-form ($C_{a_1 \dots a_6}$) potential and a term involving a potential with a mixed symmetric Young tableau diagram $$C_{a_1 \dots a_8, b} = C_{[a_1 \dots a_8], b}.$$ Except for the term involving $C_{a_1 \dots a_8, b},$ the potential is the same as that obtained in the $E_7$ non-linear realisation \cite{BGPW}:
\begin{align}
 V &= R(g) -\frac{1}{48} g^{a_1 \dots a_4, b_1 \dots b_4} F^{(4)}_{a_1 \dots a_4} F^{(4)}_{b_1 \dots b_4} - \frac{1}{8!} g^{a_1 \dots a_7, b_1 \dots b_7} F^{(7)}_{a_1 \dots a_7} F^{(7)}_{b_1 \dots b_7} \notag \\[7pt]
& \qquad \qquad\qquad \qquad\qquad \qquad \qquad\qquad  +  \frac{1}{8(8!)} g^{ad} g^{bc} g^{e_1 \dots e_8, f_1 \dots f_8} F_{a, e_1 \dots e_8, b} F_{c, f_1 \dots f_8, d},
\end{align}
where $R(g)$ is the Ricci scalar of metric $g$ and
\begin{gather}
 F^{(4)}_{a_1 \dots a_4} = 4 \partial_{[a_1} C_{a_2 a_3 a_4]},\\[5pt]
F^{(7)}_{a_1 \dots a_7} = 7 \left( \partial_{[a_1} C_{a_2 \dots a_7]} + 20 C_{[a_1 a_2 a_3} \partial_{a_4} C_{a_5 a_6 a_7]} \right), \\[5pt]
F_{a, e_1 \dots e_8, b} = \partial_{a} C_{e_1 \dots e_8, b} - 28 C_{[e_1 \dots e_6|} \partial_{a} C_{|e_7 e_8] b} - \frac{560}{3} C_{b [ e_1 e_2} C_{e_3 e_4 e_5|} \partial_a C_{|e_6 e_7 e_8]}. 
\end{gather}
The antisymmetrisation over inverse metrics is defined as follows
\begin{align}
 g^{a_1 \dots a_n, b_1 \ldots b_n}=\frac{1}{n!} & \left(g^{a_1 b_1} \dots g^{a_n b_n} + (\text{remaining even permutations of $a_1, \dots, a_n$}) \right. \notag \\
& \hspace{50mm} \left.  - \frac{}{} (\text{odd permutations of $a_1, \dots, a_n$}) \right).
\label{antisymnormg}
\end{align}
From an eleven-dimensional perspective the field strength $F^{(7)}$ is the Hodge dual of $F^{(4)}$, while the interpretation of $F_{a, b_1 \dots b_8, c}$ is unclear. Although the structure of the potential $C_{a_1 \dots a_8, b}$ suggests a relation to the dual gravity field. 

To establish such a relation, in section \ref{redsec}, we dimensionally reduce the bosonic action of eleven-dimensional supergravity to three dimensions. The reduced theory is known to exhibit $E_8$ global symmetry. Indeed upon dualising the one-forms, the scalars of the theory parametrise the coset $E_8/$SO(16) and their action is written in terms of the generalised metric. Whereupon, we identify the 6-form potential as the dual of the 3-form. Furthermore, the potential $C_{a_1 \dots a_8, b}$ is the dual of the graviphoton with field strength
$$H_{i, a_1 \dots a_8, b} = \partial_{i} C_{a_1 \dots a_8, b} - 28 C_{[a_1 \dots a_6|} \partial_{i} C_{|a_7 a_8] b} - \frac{560}{3} C_{b [a_1 a_2} C_{a_3 a_4 a_5|} \partial_i C_{|a_6 a_7 a_8]}$$
in the three-dimensional theory.

We discuss the possible implications of these results at the end of the paper.

\section{Non-linear realisation of $E_8$}
\label{genmet}

In this section, we construct the non-linear realisation of the $E_8$ motion group.\footnote{See appendix \ref{appalg} for a review of the $E_8$ algebra.} The dynamics obtained from this construction can then be compared with eleven-dimensional supergravity. Due to the many technicalities and long calculations involved in obtaining this result much of the details concerning the calculations have been explained in the appendices for ease of reading.

As emphasised before, the key ingredient in the construction of the $E_8$ invariant dynamics is the $E_8$ generalised metric, which is constructed using a non-linear realisation of the $E_8$ motion group. In \cite{BGPW}, the non-linear realisation method was used to calculate the generalised metrics relevant for the SL(5), SO(5,5), $E_6$ and $E_7$ duality groups. In that paper, the duality groups were regarded as subgroups of $E_{11}$ and the generalised metrics were found by performing a non-linear realisation of $E_{11} \ltimes l_1$ decomposed to the appropriate duality subgroup. As was stressed in that paper the only difference in carrying out the non-linear realisation of $E_{11}$ truncated to the aforementioned mentioned duality groups as opposed to doing the non-linear realisation of the duality group itself is an overall factor of the determinant of the spatial metric to some power multiplying the generalised metric. The approach that we will take in this paper is to calculate the non-linear 
realisation of the $E_8$ duality group. We comment on the overall factor of the generalised metric later in this section.

The first step in constructing the non-linear realisation is to find the $E_8$ motion group, which is done in appendix \ref{appalg}.  In order to do this, first the $E_8$ algebra, which is usually written in terms of an SL(9) decomposition of $E_8$, needs to be rewritten in terms of an SL(8) decomposition.  This is because in the 8+3 splitting of the eleven-dimensional theory that we are considering here, the $E_8$ duality group acts only on the eight spatial directions. In an SL(9) representation, the $E_8$ algebra is given by the following three generators
\begin{equation}
  M^{\underline{\alpha}}{}_{\underline{\beta}},\ V^{\underline{\alpha} \underline{\beta} \underline{\gamma}},\ V_{\underline{\alpha} \underline{\beta} \underline{\gamma}},
\end{equation}
where the underlined Greek indices are SL(9) indices that run from 1 to 9.  An SL(8) decomposition of these generators is simply of the form
\begin{equation}
 M^{\alpha}{}_{\beta},\  M^{\alpha}{}_{9},\  M^{9}{}_{\beta},\ V^{\alpha \beta \gamma} ,\ V^{\alpha \beta 9},\ V_{\alpha \beta \gamma} ,\ V_{\alpha \beta 9},
\end{equation}
where lowercase Greek indices run from 1 to 8.
The above objects and the alternating tensor $\epsilon_{\alpha_1 \ldots \alpha_8}$ in eight dimensions can then be used to define the generators of the particular SL(8) decomposition of the $E_8$ algebra used here
\begin{equation} \label{E8gens}
 K^{\alpha}{}_{\beta},\  R^{\alpha \beta \gamma} ,\ R_{\alpha \beta \gamma} ,\ R^{\alpha_1 \ldots \alpha_6},\ R_{\alpha_1 \ldots \alpha_6},\
R^{\alpha_1 \ldots \alpha_8, \beta},\ R_{\alpha_1 \ldots \alpha_8, \beta}.
\end{equation}
The precise relation between these generators and those listed above is given in appendix \ref{appalg}.  Note that as emphasised there, the definition of these generators is one particular choice out of many possibilities and has been made with the efficient calculation of the non-linear realisation in mind. Now the $E_8$ algebra reduces to a set of commutation relations involving the generators listed in \eqref{E8gens} (see equations \eqref{e8com1}--\eqref{e8com7} in appendix \ref{appalg}).

Given the SL(8) decomposition of the $E_8$ group, the next step is to define the translation generators. Consider the 248-dimensional fundamental (and adjoint) representation of $E_8$ given by generators 
\begin{equation}
  P^{\underline{\alpha}}{}_{\underline{\beta}},\ \overline{Z}^{\underline{\alpha} \underline{\beta} \underline{\gamma}},\ \overline{Z}_{\underline{\alpha} \underline{\beta} \underline{\gamma}},
\end{equation}
which are of the same form as the $E_8$ generators.  The SL(8) decomposition of the trace-free generator $P^{\underline{\alpha}}{}_{\underline{\beta}}$ gives the following generators 
\begin{equation}
 P_{\alpha},\ W^{\alpha}{}_{\beta},\ W,\ Z^{\alpha},
\end{equation}
where $W^{\alpha}{}_{\beta}$ is trace-free. A simple counting confirms that these generators have the same degrees of freedom as $P^{\underline{\alpha}}{}_{\underline{\beta}}$.  Similarly, the totally antisymmetric generators $\overline{Z}^{\underline{\alpha} \underline{\beta} \underline{\gamma}}$ and $\overline{Z}_{\underline{\alpha} \underline{\beta} \underline{\gamma}}$ are rewritten in terms of SL(8) indices as
\begin{equation}
 Z^{\alpha \beta \gamma},\ Z^{\alpha \beta}
\end{equation}
and
\begin{equation}
 W_{\alpha \beta \gamma},\ W_{\alpha \beta},
\end{equation}
respectively.  The precise definition of the above generators is given in equations \eqref{e8trans1}--\eqref{e8trans8} in appendix \ref{appalg}.  In addition, the necessary commutation relations and inner products involving the translation generators are given in appendix \ref{appalg}.

The $E_{8}$ non-linear realisation is constructed using the motion group element
$$ g = g_{l} g_{E} ,$$
where 
$$g_{E} = \e^{h_{\alpha}{}^{\beta } K^{\alpha}{}_{\beta }} \e^{\frac{1}{3!} C_{\alpha_1 \dots \alpha_3} R^{\alpha_1 \dots \alpha_3}} \e^{\frac{1}{6!} C_{\alpha_1 \dots \alpha_6} R^{\alpha_1 \dots \alpha_6}} \e^{\frac{1}{8!} C_{\alpha_1 \dots \alpha_8, \beta } R^{\alpha_1 \dots \alpha_8, \beta }}$$
is an $E_{8}$ group element. This group element has been gauge-fixed so that it mostly contains generators corresponding to negative roots, i.e.~generators of the Borel subalgebra. The only exception being the $K^{\alpha}{}_{\beta}$ which contains generators corresponding to both positive and negative roots as well as Cartan subalgebra generators. This group element introduces the fields 
$$h_{\alpha}{}^{\beta },\ C_{\alpha_1 \dots \alpha_3},\ C_{\alpha_1 \dots \alpha_6},\ C_{\alpha_1 \dots \alpha_8, \beta }.$$ 
The group element 
$$g_l = \e^{x^\alpha P_{\alpha}} \e^{ \frac{1}{\sqrt{2}} y_{\alpha \beta } Z^{\alpha \beta }} \e^{ \frac{1}{\sqrt{6}} w^{\alpha \beta \gamma } W_{\alpha \beta \gamma }} \e^{ w_{\alpha}{}^{\beta } W^{\alpha}{}_{\beta }} \e^{ 2 \sqrt{2} w W} \e^{ 4 \sqrt{\frac{2}{3}} z_{\alpha \beta \gamma } Z^{\alpha \beta \gamma }} \e^{ 4 \sqrt{2} w^{\alpha \beta } W_{\alpha \beta }} \e^{ 8 z_{\alpha} Z^{\alpha}}$$
is generated by the translation generators. The coefficient of each exponent in the group element has been chosen based on the normalisation of the translation generator, given in appendix \ref{appalg}, so that the flat metric takes the canonical form
\begin{align*}
 \d s^2 = & \delta_{\alpha \beta}\; \d x^{\alpha} \d x^{\beta} + \delta^{\alpha \beta, \gamma \delta}\; \d y_{\alpha \beta } \d y_{\gamma \delta } + \delta_{\alpha \beta \gamma, \delta \epsilon \zeta}\; \d w^{\alpha \beta \gamma } \d w^{\delta \epsilon \zeta} + \delta_{\alpha}^{\delta} \delta^{\beta}_{\gamma}\; \d w^{\alpha}{}_{\beta } \d w^{\gamma}{}_{\delta } \\[3pt]
&+ \d w \d w +\delta^{\alpha \beta \gamma, \delta \epsilon \zeta}\; \d z_{\alpha \beta \gamma } \d z_{\delta \epsilon \zeta } +\delta_{\alpha \beta, \gamma \delta}\; \d w^{\alpha \beta } \d w^{\gamma \delta } +\delta^{\alpha \beta}\; \d z_{\alpha} \d z_{\beta},
\end{align*}
where 
\begin{align}
 \delta^{\alpha_1 \ldots \alpha_n, \beta_1 \ldots \beta_n}=\frac{1}{n!} & \left(\delta^{\alpha_1 \beta_1} \ldots \delta^{\alpha_n \beta_n} + (\text{remaining even permutations of $\alpha_1, \ldots, \alpha_n$}) \right. \notag \\
& \hspace{50mm} \left.  - \frac{}{} (\text{odd permutations of $\alpha_1, \ldots, \alpha_n$}) \right).
\label{antisymnorm}
\end{align}

The generalised vielbein is given by conjugating the Maurer-Cartan form of $g_{l}$ by $g_{E}$ 
\begin{align}
P_{\Pi} \tilde{L}^{\Pi}{}_{A} \d Z^{A} = g^{-1}_{E} (g_{l}^{-1} \d g_{l}) g_{E}.
\end{align}
In this paper uppercase Greek letters denote generalised tangent space indices, while uppercase Latin indices denote generalised coordinate indices\footnote{Note that in \cite{BGPW}, opposite conventions were used for uppercase Greek and Latin indices}.

Using Hadamard's Lemma 
$$ \e^{X} Y \e^{-X} = \sum_{n=0}^{\infty} \textstyle{\frac{1}{n!}} (\textup{ad}^{n} X) Y,$$
where $$(\textup{ad} X) Y = [X, Y],$$
and the commutation relations between $K^{\alpha}{}_{\beta}$ and the translation generators, equations \eqref{comKP}--\eqref{comR81W}, we find $\tilde{L}^{\Pi}{}_{A},$ the generalised vielbein. This is a 8$\times$8 block, lower triangular matrix that is sextic in $C_{\alpha_1 \dots \alpha_3},$ cubic in $C_{\alpha_1 \dots \alpha_6}$ and quadratic in $C_{ \alpha_1 \dots \alpha_8, \beta}.$ The generalised metric is given by 
$$\tilde{M}_{AB} = \delta_{\Pi \Sigma} \tilde{L}^{\Pi}{}_{A} \tilde{L}^{\Sigma}{}_{B}.$$
However, for calculating the action, it is much easier to use the following rewriting of the generalised metric
\begin{equation}
 \tilde{M}_{AB}= G_{CD} \tilde{L}^{C}{}_{A} \tilde{L}^{D}{}_{B},
\end{equation}
where 
$$ G_{AB} = \textup{diag}(g_{ab}, g^{d_1 d_2, e_1 e_2}, g_{g_1 \dots g_3, h_1 \dots h_3}, g^{j_1 k_1} g_{j_2 k_2} - \frac{1}{8} \delta^{j_1}_{j_2} \delta^{k_1}_{k_2}, 
1, g^{m_1 \dots m_3, n_1 \dots n_3}, g_{q_1 q_2, r_1 r_2}, g^{x y}).$$ 
The index 
$$A= (a, d_1 d_2, g_1 \dots g_3, j_1 j_2, \varnothing , m_1 \dots m_3, q_1 q_2, x)$$ 
and similarly 
$$B= (b, e_1 e_2, h_1 \dots h_3, k_1 k_2, \varnothing, n_1 \dots n_3, r_1 r_2, y),$$
where $\varnothing$ denotes the fact that the corresponding object has no index.
Furthermore, 
\begin{equation}
 \tilde{L}^{A}{}_{B} = e_{\Pi}{}^{A} \tilde{L}^{\Pi}{}_{B},
\label{genvielrel}
\end{equation}
where 
\begin{align*}
 e_{\Pi}{}^{A} =& \textup{diag}(e_{\alpha}{}^{a}, e_{[d_1}{}^{\beta_1} e_{d_2]}{}^{\beta_2} , e_{\gamma_1}{}^{[g_1} \dots e_{\gamma_3}{}^{g_3]},  \\
 & \hspace{50mm}e_{j_1}{}^{\delta_1} e_{\delta_2}{}^{j_2} - \frac{1}{8} \delta^{j_2}_{j_1} \delta^{\delta_1}_{\delta_2}, 1, e_{[m_1}{}^{\epsilon_1} \dots e_{m_3]}{}^{\epsilon_3}, e_{\zeta_1}{}^{[q_1} e_{\zeta_2}{}^{q_2]}, e_{x}{}^{\eta}).
\end{align*}
$e_{a}{}^{\alpha}$ is the spatial vielbein, $$g_{ab}= \delta_{\alpha \beta} e_{a}{}^{\alpha} e_{b}{}^{\beta},$$ and $e_{\alpha}{}^{a}$ is the inverse vielbein. The index $$\Pi= (\alpha, \beta_1 \beta_2, \gamma_1 \dots \gamma_3, \delta_1 \delta_2, \varnothing, \epsilon_1 \dots \epsilon_3, \zeta_1 \zeta_2, \eta).$$

By introducing $\tilde{L}^{A}{}_{B},$ we have removed spatial vielbeine from the generalised vielbein and instead only work with the spatial metric. This is more convenient and it is the form of the generalised vielbein that will be used to calculate the action. Note that the generalised metric constructed from the $E_{8}$ motion group, $\tilde{M},$ is unit determinant. We will consider a rescaling of this generalised metric by the determinant of the spatial metric. As was explained in \cite{BGPW}---in particular appendix B---this can be thought of as considering $E_{8}$ as a subgroup of a larger group, $E_{11},$ for example. Or alternatively we can think of the SL(8) in $E_8$ as a subgroup of a larger special linear group, SL(11) for instance. This makes sense physically because the theory of course only makes sense in eleven-dimensions and we should always view the eight spatial coordinates we have here as being augmented by three other directions. The rescaled generalised vielbein that we use is $$ 
L^{A}{}_{B} = g^{-1/2} \tilde{L}^{A}{}_{B},$$ 
where $g$ is the determinant of the spatial metric. The generalised metric that we use to construct the dynamics is 
\begin{equation}
 M_{AB}= G_{CD} L^{C}{}_{A} L^{D}{}_{B} = g^{-1} \tilde{M}_{AB} . \label{Mdef}
\end{equation}
The components of $L^{A}{}_{B}$ are given in appendix \ref{appviel}.

We follow the canonical approach of \cite{davidmalcolm} to formulate the dynamics. In this approach, there is a potential and kinetic term for the fields. In a duality-invariant description both of these are given as a scalar in terms of the generalised metric. In order to find this description, consider first the potential and write a combination of terms that reduces to the Ricci scalar when the fields are independent of the generalised coordinates: 
\begin{gather*}
 V= \frac{1}{240} M^{MN} \partial_{M} M^{KL} \partial_{N} M_{KL} 
- \frac{1}{2} M^{MN} \partial_{N} M^{KL} \partial_{L} M_{MK}
- \frac{1}{496} M^{KL} \partial_{M} M^{MN} \partial_{N} M_{KL} \\[7pt]
+ \frac{23}{15(248)^2} M^{MN} (M^{KL} \partial_{M} M_{KL})( M^{RS} \partial_{N} M_{RS}).
\end{gather*}
$M^{AB}$ is the inverse of the generalised metric. When the fields are taken to only depend on the eight usual directions this expression reduces to 
\begin{align}
 V &= R(g) -\frac{1}{48} g^{a_1 \dots a_4, b_1 \dots b_4} F^{(4)}_{a_1 \dots a_4} F^{(4)}_{b_1 \dots b_4} - \frac{1}{8!} g^{a_1 \dots a_7, b_1 \dots b_7} F^{(7)}_{a_1 \dots a_7} F^{(7)}_{b_1 \dots b_7} \notag \\[7pt]
& \qquad \qquad\qquad \qquad\qquad \qquad \qquad\qquad  +  \frac{1}{8(8!)} g^{ad} g^{bc} g^{e_1 \dots e_8, f_1 \dots f_8} F_{a, e_1 \dots e_8, b} F_{c, f_1 \dots f_8, d}, \label{e8pot}
\end{align}
where $R(g)$ is the Ricci scalar of metric $g$ and
\begin{gather}
 F^{(4)}_{a_1 \dots a_4} = 4 \partial_{[a_1} C_{a_2 a_3 a_4]},\\[5pt]
F^{(7)}_{a_1 \dots a_7} = 7 \left( \partial_{[a_1} C_{a_2 \dots a_7]} + 20 C_{[a_1 a_2 a_3} \partial_{a_4} C_{a_5 a_6 a_7]} \right), \\[5pt]
F_{a, e_1 \dots e_8, b} = \partial_{a} C_{e_1 \dots e_8, b} - 28 C_{[e_1 \dots e_6|} \partial_{a} C_{|e_7 e_8] b} - \frac{560}{3} C_{b [ e_1 e_2} C_{e_3 e_4 e_5|} \partial_a C_{|e_6 e_7 e_8]}. \label{dualgravfield1}
\end{gather}
The details of this calculation are in appendix \ref{Vcal}. The kinetic term can be evaluated similarly and contains the kinetic terms associated with the metric and the 3-form \cite{davidmalcolm} and an analogous term for the 6-form. Moreover, it contains a term quadratic in time-derivatives of $C_{a_1, \dots a_8, b}.$ 

The interpretation of the appearance of a field with mixed indices, especially in the form above, in the dynamics is unclear. However, the structure of the potential is clearly reminiscent of a dual gravity field.  In the next section, we show that the potential $C_{a_1 \dots a_8, b}$ is the dual of the graviphoton in the dimensional reduction of eleven-dimensional supergravity to three dimensions.  Thus, given the evidence for the relation between dualisation of fields before and after reduction \cite{cremmerjulialupope1,CJLP2}, from an eleven-dimensional perspective this potential is indeed to be interpreted as a dual gravity field.

\section{Dimensional reduction of the bosonic sector of eleven-dimensional supergravity}
\label{redsec}

In this section, we dimensionally reduce the bosonic part of eleven-dimensional supergravity \cite{CJS} \emph{\`{a} la} Cremmer-Julia \cite{cremmerjulia} to three dimensions and relate $C_{e_1 \dots e_8, b}$ to the dual gravity field. This is the dimensional reduction in which the $E_{8}$ symmetry appears \cite{juliae8}. In particular the scalars of the reduced theory are described by an $E_{8}/$SO(16) coset, which we will demonstrate explicitly in this section. 

The bosonic part of the action of eleven-dimensional supergravity is
\begin{equation}
 S= \int \sqrt{G} \left( R(G) - \frac{1}{48} F^{ABCD} F_{ABCD} - \frac{1}{12^4} \epsilon^{A_1 \dots A_{11}} F_{A_1 \dots A_4} F_{A_5 \dots A_8} C_{A_6 A_7 A_8} \right).
 \label{11daction}
\end{equation}
Here $G$ is the eleven-dimensional metric, $C_{ABC}$ is the 3-form of eleven-dimensional supergravity and $$F_{ABCD}= 4 \partial_{[A} C_{BCD]}.$$ The index notation used in this section is different to that used elsewhere. In this section uppercase Latin letters run from 0 to 10, lowercase letters from the start of the Latin alphabet, $a, b, c, \dots,$ denote internal indices, while those from the middle of the alphabet, $ i, j, k, \dots,$ denote 3-dimensional indices. The symbol $\epsilon$ in equation \eqref{11daction}, as elsewhere in this paper, denotes an alternating tensor. 

To perform the reduction, we take all fields to be independent of the internal directions. First, consider the gravitational part. We take the following ansatz for the elfbein
\begin{equation}
 \begin{pmatrix}
g^{-1/2} e_{i}{}^{\mu} & B_{i}{}^{a} \tilde{e}_{a}{}^{\alpha} \\
0 & \tilde{e}_{a}{}^{\alpha} 
 \end{pmatrix},
 \label{vielbeinansatz}
\end{equation}
where $e_{i}{}^{\mu}$ and $\tilde{e}_{a}{}^{\alpha}$ are the dreibein and achtbein. In this section, lowercase Greek indices from the beginning and middle of the alphabet denote internal and 3-dimensional tangent space indices, respectively. We define the three-dimensional and eight-dimensional metrics as follows:
\begin{align}
 \gamma_{ij} &= e_{i}{}^{\mu} e_{j}{}^{\nu} \eta_{\mu \nu} \label{3met}\\
 \textup{and} \qquad  g_{ab} &= \tilde{e}_{a}{}^{\alpha} \tilde{e}_{b}{}^{\beta} \delta_{\alpha \beta}, \label{7met}
\end{align}
so that $g$ in expression \eqref{vielbeinansatz} denotes the determinant of metric $g_{ab}.$ 
Given the vielbein ansatz \eqref{vielbeinansatz}, 
\begin{equation}
 \sqrt{G} R(G) = \sqrt{\gamma} \left( R(\gamma) + \frac{1}{4} \gamma^{ij} (\partial_{i} g^{ab})( \partial_{j} g_{ab} ) - \frac{1}{4} \gamma^{ij} (g^{ab} \partial_{i} g_{ab}) (g^{cd} \partial_{j} g_{cd}) - \frac{1}{4} g  \gamma^{i k} \gamma^{jl} g_{ab} F_{ij}{}^{a} F_{kl}{}^{b} \right),
 \label{redgrav}
\end{equation}
where $$F_{ij}{}^{a} = 2 \partial_{[i} B_{j]}{}^{a} $$
is the field strength of the graviphoton.  

Under the reduction, the second term in the action, given in \eqref{11daction}, becomes 
\begin{equation}
- \frac{1}{48} \sqrt{G} F^{ABCD} F_{ABCD} = \sqrt{\gamma} \left( - \frac{1}{12}  g^2 \tilde{F}^{ijka} \tilde{F}_{ijka} - \frac{1}{8} g \tilde{F}^{ijab} \tilde{F}_{ijab} - \frac{1}{12} F^{i abc} F_{iabc} \right).\label{redkin}  
\end{equation}
In the above expression the indices are raised with inverses of the metrics $\gamma$ and $g$ defined in equations \eqref{3met} and \eqref{7met}, so for example
$$\tilde{F}^{ijab} = \gamma^{ik} \gamma^{jl} g^{ac} g^{bd} \tilde{F}_{klcd}.$$ Moreover, the field strengths 
\begin{align}
 \tilde{F}_{ijkc} &= 3 \partial_{[i} C_{jk]c} - 6 (\partial_{[i} C_{j|bc}) B_{|k]}{}^{b} + 3 (\partial_{[i|} C_{abc}) B_{|j}{}^{a} B_{k]}{}^{b}, \\[3pt]
  \tilde{F}_{ijbc} &= 2 \partial_{[i} C_{j]bc} - 2 \partial_{[i|} C_{abc} B_{|j]}{}^{a}, \label{Fijabdef}\\[3pt]
  F_{iabc} &= \partial_{i} C_{abc}
\end{align}
are defined so that they are invariant under coordinate transformations of the internal directions---see \cite{cremmerjulia} for more details. 

Similarly, in terms of the gauge-invariant field strengths defined above, the Chern-Simons term of the action reduces to
\begin{equation}
 -\frac{2}{12^3} \sqrt{\gamma} \sqrt{g} \epsilon^{ijk} \epsilon^{a_1 \dots a_8} \left(3 \tilde{F}_{ija_1 a_2} F_{k a_3 a_4 a_5} - F_{i a_1 a_2 a_3} C_{a_4 a_5 b} F_{jk}{}^{b} \right) C_{a_6 a_7 a_8}. \label{redCS}
\end{equation}
In obtaining the above result we have integrated by parts twice and used $$F_{i[a_1 a_2 a_3|} F_{j| a_4 a_5 a_6} C_{a_7 a_8 a_9]}=0.$$  

Putting together equations \eqref{redgrav}, \eqref{redkin} and \eqref{redCS}, we obtain the action for the reduced theory
\begin{align}
 S^{(3)} &= \int \sqrt{\gamma} \left( R(\gamma) + \frac{1}{4} \gamma^{ij} (\partial_{i} g^{ab})( \partial_{j} g_{ab} ) - \frac{1}{4} \gamma^{ij} (g^{ab} \partial_{i} g_{ab}) (g^{cd} \partial_{j} g_{cd}) - \frac{1}{4} g  \gamma^{i k} \gamma^{jl} g_{ab} F_{ij}{}^{a} F_{kl}{}^{b} \right. \notag \\[4pt]
 & \qquad \qquad \qquad  - \frac{1}{12}  g^2 \tilde{F}^{ijka} \tilde{F}_{ijka} - \frac{1}{8} g \tilde{F}^{ijab} \tilde{F}_{ijab} - \frac{1}{12} F^{i abc} F_{iabc} \notag \\[4pt]
  & \qquad \qquad \qquad \qquad \left. -\frac{2}{12^3} \sqrt{g} \epsilon^{ijk} \epsilon^{a_1 \dots a_8} \left( 3 \tilde{F}_{ija_1 a_2} F_{k a_3 a_4 a_5} - F_{i a_1 a_2 a_3} C_{a_4 a_5 b} F_{jk}{}^{b} \right) C_{a_6 a_7 a_8} \right).
\end{align}
We are interested in the scalars of the reduced theory because it is these that parametrise the $E_{8}/$SO(16) coset. From the above action we can see that the scalars of the theory are 36 $g_{ab},$ 56 $C_{abc}.$ Furthermore, since we are in three dimensions, one-forms are dual to scalars so we have 28 + 8 scalars from dualising the one-forms $A_{iab}$ and $B_{i}{}^{a}.$ Therefore, in all we have $$128 = 248 -120= \textup{dim}(E_{8}) - \textup{dim(SO(16))}$$ scalars. We concentrate on the action of the scalars and one-forms and augment the action by a Lagrange multiplier that imposes the closedness of the one-form field strengths. 
\begin{align}
 S'^{(3)} &= \int \sqrt{\gamma} \left\{ \frac{1}{4} \gamma^{ij} (\partial_{i} g^{ab})( \partial_{j} g_{ab} ) - \frac{1}{4} \gamma^{ij} (g^{ab} \partial_{i} g_{ab}) (g^{cd} \partial_{j} g_{cd})  \right. \notag \\[4pt]
 & \qquad \qquad \qquad - \frac{1}{4} g  \gamma^{i k} \gamma^{jl} g_{ab} F_{ij}{}^{a} F_{kl}{}^{b} - \frac{1}{8} g \tilde{F}^{ijab} \tilde{F}_{ijab} - \frac{1}{12} F^{i abc} F_{iabc} \notag \\[4pt]
  & \qquad \qquad \qquad \qquad  -\frac{2}{12^3} \sqrt{g} \epsilon^{ijk} \epsilon^{a_1 \dots a_8} \left( 3 \tilde{F}_{ija_1 a_2} F_{k a_3 a_4 a_5} - F_{i a_1 a_2 a_3} C_{a_4 a_5 b} F_{jk}{}^{b} \right) C_{a_6 a_7 a_8} \notag \\[4pt]
  & \qquad \qquad \qquad \qquad \qquad \left. - \frac{1}{4} \varphi_{a} \epsilon^{ijk} \partial_{i} F_{jk}{}^{a} +\frac{1}{8} \psi^{ab} \epsilon^{ijk} \left(  \partial_{i} \tilde{F}_{jkab} - 2 F_{iabc} F_{jk}{}^{c} \right)  \right\}.
\end{align}
On a three-dimensional manifold with trivial homology, integrating out $\varphi_{a}$ gives
$$F_{ij}{}^{a}= 2 \partial_{[i} B_{j]}{}^{a}$$ 
for some $B.$ While the equation of motion for the second Lagrange multiplier, $\psi,$ gives that
$$\tilde{F}_{jkab} + 2 F_{jabc} B_{k}{}^{c}$$ 
is closed, from which we recover equation \eqref{Fijabdef}.
Therefore, this first-order formulation is, at least classically, equivalent to the action for the scalars and one-form of the original reduced action $S^{(3)}.$ Hence, $\tilde{F}_{ikab}$ and $F_{ij}{}^{a}$ are independent fields not given in terms of potential forms. By integrating out these fields we dualise the one-forms of the original action, $B_{i}^{a}$ and $C_{iab},$ into scalars $\varphi_{a}$ and $\psi^{ab}.$ In fact, this is the reason why these duality symmetries are sometimes called hidden symmetries. The symmetry is only manifest after dualisation of some of the fields. The new action that we obtain is
\begin{align}
 S^{(3)}_{\textup{scalars}} &= \int \sqrt{\gamma} \left( \frac{1}{4} \gamma^{ij} (\partial_{i} g^{ab})( \partial_{j} g_{ab} ) - \frac{1}{4} \gamma^{ij} (g^{ab} \partial_{i} g_{ab}) (g^{cd} \partial_{j} g_{cd})  \right. \notag \\[3pt]
 & \qquad \qquad \qquad \left. - \frac{1}{12} \gamma^{ij} g^{abc, def} \partial_{i} C_{abc} \partial_{j} C_{def}  - \frac{1}{2} \gamma^{ij} g^{ab} G_{i a} G_{j b}- \frac{1}{16}\gamma^{ij} g_{ab,cd} G_{i}{}^{ab} G_{j}{}^{cd} \right),
\end{align}
where, $g^{a_1 \dots a_n, b_1 \dots b_n}$ is defined as in equation \eqref{antisymnormg}. $g_{a_1 \dots a_n, b_1 \dots b_n}$ is defined analogously.  The new fields in the action are
\begin{align}
  G_{i}{}^{ab} &= g^{-1/2} \partial_{i} \psi^{ab} - \frac{1}{36} \epsilon^{ab c_1 \dots c_6} C_{c_1 c_2 c_3} \partial_{i} C_{c_4 c_5 c_6}, \\[3pt]
  G_{i a} &= g^{-1/2} \partial_{i} \varphi_{a} - 1/2 g^{-1/2} \psi^{bc} \partial_{i} C_{abc} - \frac{1}{216} \epsilon^{b_1 \dots b_8} C_{a b_1 b_2} C_{b_3 b_4 b_5} \partial_i C_{b_6 b_7 b_8}.
\end{align}
Defining 
\begin{align}
 C_{a_1 \dots a_6} &= \frac{1}{2} g^{-1/2} \epsilon_{a_1 \dots a_6 bc} \psi^{bc},\\[2pt] 
 C_{a_1 \dots a_8, b} &= g^{-1/2} \epsilon_{a_1 \dots a_8} \varphi_{b}
\end{align}
we can identify these fields with the dual of the 3-form and the dual gravity field. With this notation for the fields the action of the scalars in three dimensions can be written   
\begin{align}
 S^{(3)}_{\textup{scalars}} &= \int \sqrt{\gamma} \left( \frac{1}{4} \gamma^{ij} (\partial_{i} g^{ab})( \partial_{j} g_{ab} ) - \frac{1}{4} \gamma^{ij} (g^{ab} \partial_{i} g_{ab}) (g^{cd} \partial_{j} g_{cd})  \right. \notag \\[3pt]
 & \qquad \qquad \qquad  - \frac{1}{12} \gamma^{ij} g^{abc, def} \partial_{i} C_{abc} \partial_{j} C_{def}  - \frac{1}{8(6!)} \gamma^{ij} g^{a_1 \dots a_6, b_1 \dots b_6} H_{i, a_1 \dots a_6} H_{j, a_1 \dots a_6} \notag \\[3pt]
 & \qquad \qquad \qquad \qquad \left.- \frac{1}{8(8!)} \gamma^{ij} g^{a_1 \dots a_8, b_1 \dots b_8} g^{cd} H_{i, a_1 \dots a_8, c} H_{j, a_1 \dots a_8, c} \right), \label{3dscalaraction}
\end{align}
where 
\begin{gather}
H_{i, a_1 \dots a_6} =  \partial_{i} C_{a_1 \dots a_6} - 20 C_{[a_1 a_2 a_3|} \partial_{i} C_{|a_4 a_5 a_6]}, \\[4pt]
H_{i, a_1 \dots a_8, b} = \partial_{i} C_{a_1 \dots a_8, b} - 28 C_{[a_1 \dots a_6|} \partial_{i} C_{|a_7 a_8] b} - \frac{560}{3} C_{b [a_1 a_2} C_{a_3 a_4 a_5|} \partial_i C_{|a_6 a_7 a_8]}. \label{dualgravfield2}
\end{gather}
As expected, since the scalars in the reduction parametrise the $E_{8}/$SO(16) coset, action \eqref{3dscalaraction} can be written in terms of the $E_8$ generalised metric, \eqref{Mdef}, in the following way
\begin{equation}
 S^{(3)}_{\textup{scalars}} = \frac{1}{240} \gamma^{ij} \partial_{i} M^{KL} \partial_{j} M_{KL} 
+ \frac{31}{30(248)^2} \gamma^{ij} (M^{KL} \partial_{i} M_{KL})( M^{RS} \partial_{j} M_{RS}),
\end{equation}
where the uppercase Latin indices in the above equation run from 1 to 248 as in section \ref{genmet}. Note that the calculation of the above terms is identical to the calculation of the potential in appendix \ref{Vcal}. 

Comparing equations \eqref{dualgravfield2} and \eqref{dualgravfield1}, we can see that it is the dual gravity field that appears in the potential in section \ref{genmet}. This is in contrast to the $E_6$ case considered in reference \cite{BGPW} where the 6-form field could have appeared in the potential but didn't because there was an antisymmetrisation over 7 indices. In the potential given in equation \eqref{e8pot} there is no antisymmetrisation over the first 9 indices of $F_{a, b_1 \dots b_8, c}$ so the dual gravity field appears.  From a technical viewpoint, this is because $C_{a_1 \ldots a_8, b}$ has mixed indices.  Note that, in contrast to the field strength $F_{a, e_1 \dots e_8, b}$ defined in equation \eqref{dualgravfield1}, the gauge invariance of $H_{i, a_1 \dots a_8, b}$ from a three-dimensional point-of-view is very clear to see.  This is because in the reduced theory, the fields $C_{abc}$, $C_{a_1, \ldots a_6}$ and $C_{a_1 \ldots a_8, b}$ are scalars. 

\section{Discussion}

In this paper, we formulated a non-linear realisation of the $E_8$ group and found that the dynamics includes a new field $C_{a_1 \ldots a_8, b}$ with mixed Young tableaux indices with field strength $F_{a, b_1 \dots b_8, c}$.  While the gauge-invariance properties of $F_{a, b_1 \dots b_8, c}$ are not clear, we show tantalising links with dual gravity.  Our difficulty in establishing the gauge-invariance of the field strength is related to the difficulty in formulating a generalised geometry for $E_8$ \cite{CSWex, BCKT}.  In both cases, knowledge of the transformation of $C_{a_1 \ldots a_8, b}$ under gauge transformations is a requisite.

In reference \cite{BCKT}, the authors were unable to write down a generalised Lie derivative, even though they showed that the gauge structure leads to the correct counting of the degrees of freedom.  While, as in \cite{BCKT}, we cannot determine the gauge transformations of $C_{a_1 \ldots a_8, b}$, we unambiguously show that if the field strength is to be gauge-invariant, the new field must transform under 3-form and 6-form gauge transformations.  This result establishes the possible dependence of the dual gravity field on the eleven-dimensional matter fields, namely the 3-form gauge field and its dual and may provide a basis for evading the no-go theorems of \cite{BBH, BdRKKR}.  In reference \cite{BdRKKR}, it was shown that even a linearised dual gravity formulation is not possible in the presence of matter unless covariance or locality is abandoned\footnote{We thank Axel Kleinschmidt for discussions on this point.}. It is possible that $C_{a_1 \ldots a_8, b}$ is dual to a particular component of the eleven-dimensional metric, which is consistent with the reduced theory perspective. We leave a precise description of such a possibility for future work.  

\medskip

\paragraph*{Acknowledgements}

We would like to thank David Berman, Chris Blair, Axel Kleinschmidt and Hermann Nicolai for discussions. HG would like to thank the CERN theory group, where part of this work was done, for their hospitality. MJP is in part supported by the STFC rolling grant STJ000434/1. MJP would like to thank the Mitchell foundation and Trinity College Cambridge for their generous support.

\newpage
\appendix

\section{$E_8$ motion group from Cartan's representation}
\label{appalg}

In this appendix, we find the algebra of the $E_{8}$ motion group, where the translation generators of the motion group form the 248-dimensional representation of $E_{8}.$ In particular, the algebra of the $E_{8}$ motion group decomposed to SL(8) is found. 

The $E_{8}$ \cite{cartan} group is generated by
\begin{equation}
  M^{\underline{\alpha}}{}_{\underline{\beta}}, V^{\underline{\alpha} \underline{\beta} \underline{\gamma}}, V_{\underline{\alpha} \underline{\beta} \underline{\gamma}},
\label{e8gen}
\end{equation}
where underlined Greek indices run from 1 to 9. In terms of these generators the $E_{8}$ algebra is as follows \cite{cartan}:
\begin{gather}
 [ M^{\underline{\alpha}}{}_{\underline{\beta}}, M^{\underline{\gamma}}{}_{\underline{\delta}} ] = \delta^{\underline{\gamma}}_{\underline{\beta}} M^{\underline{\alpha}}{}_{\underline{\delta}} 
- \delta^{\underline{\alpha}}_{\underline{\delta}} M^{\underline{\gamma}}{}_{\underline{\beta}}, \label{e8alg:sl91}\\
[M^{\underline{\alpha}}{}_{\underline{\beta}}, V^{ \underline{\gamma}_{1} \dots \underline{\gamma}_{3}}] = 3 \delta_{\underline{\beta}}^{[\underline{\gamma}_{1}} V^{ \underline{\gamma}_{2} \underline{\gamma}_{3}] \underline{\alpha}} -\frac{1}{3} \delta^{\underline{\alpha}}_{\underline{\beta}} V^{\underline{\gamma}_{1} \dots \underline{\gamma}_{3}}, \\
[M^{\underline{\alpha}}{}_{\underline{\beta}}, V_{ \underline{\gamma}_{1} \dots \underline{\gamma}_{3}}] = - 3 \delta^{\underline{\alpha}}_{[\underline{\gamma}_{1}} V_{ \underline{\gamma}_{2} \underline{\gamma}_{3}] \underline{\beta}} +\frac{1}{3} \delta^{\underline{\alpha}}_{\underline{\beta}} V_{\underline{\gamma}_{1} \dots \underline{\gamma}_{3}},\\
[V^{\underline{\alpha}_{1} \dots \underline{\alpha}_3}, V_{\underline{\beta}_1 \dots \underline{\beta}_3}] = 18 \delta^{[\underline{\alpha}_1 \underline{\alpha}_2}_{[\underline{\beta}_1 \underline{\beta}_2} M^{\underline{\alpha}_3]}{}_{\underline{\beta}_3]}, \\
[V^{\underline{\alpha}_1 \dots \underline{\alpha}_3}, V^{\underline{\beta}_1 \dots \underline{\beta}_3}] = - \frac{1}{3!} \epsilon^{ \underline{\alpha}_1 \underline{\alpha}_2 \underline{\alpha}_3 \underline{\beta}_1 \underline{\beta}_2 \underline{\beta}_3 \underline{\gamma}_1 \underline{\gamma}_2 \underline{\gamma}_3 } V_{ \underline{\gamma}_1 \dots \underline{\gamma}_3},\\
[V_{\underline{\alpha}_1 \dots \underline{\alpha}_3}, V_{\underline{\beta}_1 \dots \underline{\beta}_3}] = \frac{1}{3!} \epsilon_{ \underline{\alpha}_1 \underline{\alpha}_2 \underline{\alpha}_3 \underline{\beta}_1 \underline{\beta}_2 \underline{\beta}_3 \underline{\gamma}_1 \underline{\gamma}_2 \underline{\gamma}_3 } V^{ \underline{\gamma}_1 \dots \underline{\gamma}_3},
\label{e8alg:sl96}
\end{gather}
where $\delta^{\underline{\alpha}}_{\underline{\beta}}$ is the Kronecker delta symbol and $\epsilon_{\underline{\alpha}_1 \dots \underline{\alpha}_9}$ is the alternating tensor in nine dimensions. Furthermore,
$$\epsilon^{\underline{\alpha}_1 \dots \underline{\alpha}_9} \epsilon_{\underline{\beta}_1 \dots \underline{\beta}_9} = 9! \; \delta^{\underline{\alpha}_1 \dots \underline{\alpha}_9}_{\underline{\beta}_1 \dots \underline{\beta}_9}.$$ The $E_8$ algebra is expressed in terms of an SL(9) decomposition of $E_{8}.$ In this paper, we are considering the action of the $E_8$ duality group along eight dimensions. Hence we require an SL(8) decomposition of the algebra. This is easily done by defining $E_{8}$ generators in SL(8) representations as follows
\begin{gather}    
K^{\alpha }{}_{\beta } = M^{\alpha }{}_{\beta } + \delta^{\alpha }_{\beta } \sum_{\gamma =1}^{8} M^{\gamma }{}_{\gamma }, \\
R^{\alpha  \beta  \gamma }= V^{\alpha \beta \gamma }, \\
R_{\alpha  \beta  \gamma }= V_{\alpha \beta \gamma }, \\
R^{\alpha _1 \dots \alpha _6} = - \frac{1}{4} \epsilon^{\alpha _1 \dots \alpha _6 \beta \gamma} V_{\beta \gamma  9}, \\
R_{\alpha _1 \dots \alpha _6} = \frac{1}{4} \epsilon_{\alpha _1 \dots \alpha _6 \beta \gamma } V^{\beta \gamma  9}, \\
R^{\alpha _1 \dots \alpha _8, \beta } = \frac{1}{2} \epsilon^{\alpha _1 \dots \alpha _8} M^{\beta }{}_{9}, \\
R_{\alpha _1 \dots \alpha _8, \beta } = \frac{1}{2} \epsilon_{\alpha _1 \dots \alpha _8} M^{9}{}_{\beta },
\end{gather}
where Greek indices are SL(8) indices. The alternating tensor in eight dimensions is induced from the nine-dimensional one in the following way:
$$\epsilon_{\alpha _1 \dots \alpha _8}= \epsilon_{\alpha _1 \dots \alpha _8 9} \qquad \textup{and} \qquad \epsilon^{\alpha _1 \dots \alpha _8}= \epsilon^{\alpha _1 \dots \alpha _8 9}. $$
Using equations \eqref{e8alg:sl91}--\eqref{e8alg:sl96} and the above equations, we find the $E_{8}$ algebra given in terms of an SL(8) decomposition. The commutation relations of the GL(8) generator are 
\begin{gather}
[K^{\alpha  }{}_{\beta }, K^{\gamma }{}_{\delta}] = \delta^{\gamma }_{\beta } K^{\alpha  }{}_{\delta} - \delta^{\alpha  }_{\delta} K^{\gamma }{}_{\beta }, \label{e8com1} \\
[K^{\alpha  }{}_{\beta }, R^{\gamma _1 \dots \gamma _3}] = 3 \delta^{[\gamma _1}_{\beta } R^{|\alpha  | \gamma _2 \gamma _3]}, \\
[K^{\alpha  }{}_{\beta }, R_{\gamma _1 \dots \gamma _3}] = - 3 \delta_{[\gamma _1}^{\alpha  } R_{|\beta | \gamma _2 \gamma _3]}, \\
[K^{\alpha  }{}_{\beta }, R^{\gamma _1 \dots \gamma _6}] = 6 \delta^{[\gamma _1}_{\beta } R^{|\alpha  | \gamma _2 \dots \gamma _6]}, \\
[K^{\alpha  }{}_{\beta }, R_{\gamma _1 \dots \gamma _6}] = - 6 \delta_{[\gamma _1}^{\alpha  } R_{|\beta | \gamma _2 \dots \gamma _6]}, \\
[K^{\alpha  }{}_{\beta }, R^{\gamma _1 \dots \gamma _8, \delta}] = 8 \delta^{[\gamma _1}_{\beta } R^{|\alpha  | \gamma _2 \dots \gamma _8], \delta} + \delta^{\delta}_{\beta } R^{\gamma _1 \dots \gamma _8, \alpha  }, \\
[K^{\alpha  }{}_{\beta }, R_{\gamma _1 \dots \gamma _8, \delta}] = - 8 \delta_{[\gamma _1}^{\alpha  } R_{|\beta | \gamma _2 \dots \gamma _8], \delta} - \delta_{\delta}^{\alpha  } R_{\gamma _1 \dots \gamma _8, \beta }.
\end{gather}
These are the expected commutation relations of the GL(8) generator $K^{\alpha}{}_{\beta}$ with the other generators. The generator $K^{\alpha}{}_{\beta}$ has been shifted by $\sum_{\gamma} M^{\gamma}{}_{\gamma}$ in such a way that its commutation relations with the $R$ generators do not contain any $\delta^{\alpha}{}_{\beta}.$ Other choices can be made, but this choice is more convenient and makes the non-linear realisation calculation easier. Furthermore, with this choice the trace of $K^{\alpha}{}_{\beta},$ $$K=\sum_{\gamma} K^{\gamma}{}_{\gamma},$$ counts the index of the GL(8) representations 
\begin{gather*}
[K, R^{\alpha_1 \dots \alpha_3}] = 3  R^{\alpha_1 \dots \alpha_3}, \\
[K, R_{\alpha_1 \dots \alpha_3}] = - 3 R_{\alpha_1 \dots \alpha_3}, \\
[K, R^{\alpha_1 \dots \alpha_6}] = 6  R^{\alpha_1 \dots \alpha_6}, \\
[K, R_{\alpha_1 \dots \alpha_6}] = - 6 R_{\alpha_1\dots \alpha_6}, \\
[K, R^{\alpha_1 \dots \alpha_8, \beta }] =  9 R^{\alpha_1 \dots \alpha_8, \beta }, \\
[K, R_{\alpha_1 \dots \alpha_8, \beta }] = - 9 R_{\alpha_1 \dots \alpha_8, \beta }.
\end{gather*}
The rest of the commutations relations in the SL(8) decomposition are
\begin{gather}
 [R^{\alpha_1 \dots \alpha_3}, R^{\beta _1 \dots \beta _3}] = 2 R^{\alpha_1 \alpha_2 \alpha_3 \beta _1 \beta _2 \beta _3},\\
 [R^{\alpha_1 \dots \alpha_3}, R^{\beta _1 \dots \beta _6}] = -3 R^{\beta _1 \dots \beta _6 [\alpha_1 \alpha_2, \alpha_3] },\\
 [R^{\alpha_1 \dots \alpha_3}, R_{\beta _1 \dots \beta _3}] = 18 \delta^{[\alpha_1 \alpha_2}{}_{[\beta _1 \beta _2} K^{\alpha_3]}{}_{\beta 3]} - 2 \delta^{\alpha_1 \dots \alpha_3}_{\beta _1 \dots \beta _3} K, \label{crR3uR3d}\\
 [R^{\alpha_1 \dots \alpha_3}, R_{\beta _1 \dots \beta _6}] = 60 \delta^{\alpha_1 \dots \alpha_3}_{[\beta _1 \dots \beta _3} R_{\beta _4 \dots \beta _6]},\\
 [R^{\alpha_1 \dots \alpha_3}, R_{\beta _1 \dots \beta _8, \gamma }] = 112 \left( \delta^{\alpha_1 \dots \alpha_3}_{[\beta _1 \dots \beta _3} R_{\beta _4 \dots \beta _8] \gamma } - \delta^{\alpha_1 \alpha_2 \alpha_3}_{\gamma  [\beta _1 \beta _2} R_{\beta _3 \dots \beta _8]} \right),\\
[R^{\alpha_1 \dots \alpha_6}, R_{\beta _1 \dots \beta _3}] = -60 \delta^{[\alpha_1 \dots \alpha_3}_{\beta _1 \dots \beta _3} R^{\alpha_4 \dots \alpha_6]},\\
 [R^{\alpha_1 \dots \alpha_6}, R_{\beta _1 \dots \beta _6}] = - 9 (5!) \delta^{[\alpha_1 \dots \alpha_5}_{[\beta _1 \dots \beta _5} K^{\alpha_6]}{}_{\beta 6]} + 5! \delta^{\alpha_1 \dots \alpha_6}_{\beta _1 \dots \beta _6} K,\\
 [R^{\alpha_1 \dots \alpha_6}, R_{\beta _1 \dots \beta _8, \gamma }] =  \frac{2}{3} 7!\left( \delta^{\alpha_1 \alpha_2 \dots \alpha_6}_{\gamma  [\beta _1 \dots \beta _5} R_{\beta _6 \dots \beta _8]} - \delta^{\alpha_1 \dots \alpha_6}_{[\beta _1 \dots \beta _6} R_{\beta _7 \beta _8] \gamma } \right),\\
[R^{\alpha_1 \dots \alpha_8, \beta }, R_{\gamma _1 \dots \gamma _3}] = - 112 \left( \delta^{[\alpha_1 \dots \alpha_3}_{\gamma _1 \dots \gamma _3} R^{\alpha_4 \dots \alpha_8] \beta } - \delta^{\beta  [\alpha_1 \alpha_2}_{\gamma _1 \gamma _2 \gamma _3} R^{\alpha_3 \dots \alpha_8]} \right),\\
 [R^{\alpha_1 \dots \alpha_8, \beta }, R_{\gamma _1 \dots \gamma _6}] = - \frac{2}{3} 7! \left( \delta^{\beta  [\alpha_1 \dots \alpha_5}_{\gamma _1 \gamma _2 \dots \gamma _6} R^{\alpha_6 \dots \alpha_8]} - \delta^{[\alpha_1 \dots \alpha_6}_{\gamma _1 \dots \gamma _6} R^{\alpha_7 \alpha_8] \beta } \right),\\
 [R^{\alpha_1 \dots \alpha_8, \beta }, R_{\gamma _1 \dots \gamma _8, \delta}] =  \frac{8!}{4} \delta^{\alpha_1 \dots \alpha_8}_{\gamma _1 \dots \gamma _8} K^{\beta }{}_{\delta},\\
[R_{\alpha_1 \dots \alpha_3}, R_{\beta _1 \dots \beta _3}] = 2 R_{\alpha_1 \alpha_2 \alpha_3 \beta _1 \beta _2 \beta _3},\\
 [R_{\alpha_1 \dots \alpha_3}, R_{\beta _1 \dots \beta _6}] = -3 R_{\beta _1 \dots \beta _6 [\alpha_1 \alpha_2, \alpha_3] }. \label{e8com7}
\end{gather}
All other commutation relations vanish. 

Since we are interested in the motion group of $E_{8}$ with the translation generators in the adjoint representation of $E_8,$ it is straightforward to find the algebra of the motion group from the $E_{8}$ algebra. Define translation generators $$P^{\underline{\alpha}}{}_{\underline{\beta}}, \overline{Z}^{\underline{\alpha} \underline{\beta} \underline{\gamma}}, \overline{Z}_{\underline{\alpha} \underline{\beta} \underline{\gamma}}.$$ Note that these are of the same form as $E_{8}$ generators, \eqref{e8gen}, and they satisfy analogous commutation relations with the $E_8$ generators
\begin{gather}
[ M^{\underline{\alpha}}{}_{\underline{\beta}}, P^{\underline{\gamma}}{}_{\underline{\delta}} ] = \delta^{\underline{\gamma}}_{\underline{\beta}} P^{\underline{\alpha}}{}_{\underline{\delta}} 
- \delta^{\underline{\alpha}}_{\underline{\delta}} P^{\underline{\gamma}}{}_{\underline{\beta}}, \label{e8mtnalg91} \\
[M^{\underline{\alpha}}{}_{\underline{\beta}}, \overline{Z}^{ \underline{\gamma}_{1} \dots \underline{\gamma}_{3}}] = 3 \delta_{\underline{\beta}}^{[\underline{\gamma}_{1}} \overline{Z}^{ \underline{\gamma}_{2} \underline{\gamma}_{3}] \underline{\alpha}} -\frac{1}{3} \delta^{\underline{\alpha}}_{\underline{\beta}} \overline{Z}^{\underline{\gamma}_{1} \dots \underline{\gamma}_{3}}, \\
[M^{\underline{\alpha}}{}_{\underline{\beta}}, \overline{Z}_{ \underline{\gamma}_{1} \dots \underline{\gamma}_{3}}] = - 3 \delta^{\underline{\alpha}}_{[\underline{\gamma}_{1}} \overline{Z}_{ \underline{\gamma}_{2} \underline{\gamma}_{3}] \underline{\beta}} +\frac{1}{3} \delta^{\underline{\alpha}}_{\underline{\beta}} \overline{Z}_{\underline{\gamma}_{1} \dots \underline{\gamma}_{3}},\\
[V^{ \underline{\alpha}_{1} \dots \underline{\alpha}_{3}}, P^{\underline{\beta}}{}_{\underline{\gamma}}] = - 3 \delta_{\underline{\gamma}}^{[\underline{\alpha}_{1}} \overline{Z}^{ \underline{\alpha}_{2} \underline{\alpha}_{3}] \underline{\beta}} +\frac{1}{3} \delta^{\underline{\beta}}_{\underline{\gamma}} \overline{Z}^{\underline{\alpha}_{1} \dots \underline{\alpha}_{3}}, \\
[V^{\underline{\alpha}_1 \dots \underline{\alpha}_3}, \overline{Z}^{\underline{\beta}_1 \dots \underline{\beta}_3}] = - \frac{1}{3!} \epsilon^{ \underline{\alpha}_1 \underline{\alpha}_2 \underline{\alpha}_3 \underline{\beta}_1 \underline{\beta}_2 \underline{\beta}_3 \underline{\gamma}_1 \underline{\gamma}_2 \underline{\gamma}_3 } \overline{Z}_{ \underline{\gamma}_1 \dots \underline{\gamma}_3},\\
[V^{\underline{\alpha}_{1} \dots \underline{\alpha}_3}, \overline{Z}_{\underline{\beta}_1 \dots \underline{\beta}_3}] = 18 \delta^{[\underline{\alpha}_1 \underline{\alpha}_2}_{[\underline{\beta}_1 \underline{\beta}_2} P^{\underline{\alpha}_3]}{}_{\underline{\beta}_3]}, \\
[V_{ \underline{\alpha}_{1} \dots \underline{\alpha}_{3}}, P^{\underline{\beta}}{}_{\underline{\gamma}}] =  3 \delta^{\underline{\beta}}_{[\underline{\alpha}_{1}} \overline{Z}_{ \underline{\alpha}_{2} \underline{\alpha}_{3}] \underline{\gamma}} - \frac{1}{3} \delta^{\underline{\beta}}_{\underline{\gamma}} \overline{Z}_{\underline{\alpha}_{1} \dots \underline{\alpha}_{3}}, \\
[V_{\underline{\alpha}_{1} \dots \underline{\alpha}_3}, \overline{Z}^{\underline{\beta}_1 \dots \underline{\beta}_3}] = - 18 \delta_{[\underline{\alpha}_1 \underline{\alpha}_2}^{[\underline{\beta}_1 \underline{\beta}_2} P^{\underline{\beta}_3]}{}_{\underline{\alpha}_3]}, \\
[V_{\underline{\alpha}_1 \dots \underline{\alpha}_3}, \overline{Z}_{\underline{\beta}_1 \dots \underline{\beta}_3}] = \frac{1}{3!} \epsilon_{ \underline{\alpha}_1 \underline{\alpha}_2 \underline{\alpha}_3 \underline{\beta}_1 \underline{\beta}_2 \underline{\beta}_3 \underline{\gamma}_1 \underline{\gamma}_2 \underline{\gamma}_3 } \overline{Z}^{ \underline{\gamma}_1 \dots \underline{\gamma}_3}, \label{e8mtnalg99}
\end{gather}
We similarly decompose the translation generators into an SL(8) decomposition:
\begin{gather} 
P_{\alpha } = P^{9}{}_{\alpha }, \label{e8trans1} \\
Z^{\alpha  \beta } = - \overline{Z}^{\alpha \beta 9}, \\
W_{\alpha \beta \gamma } = - \overline{Z}_{\alpha \beta \gamma }, \\
W^{\alpha }{}_{\beta } = - P^{\alpha }{}_{\beta } + \frac{1}{8} \delta^{\alpha }_{\beta } \sum_{\gamma } P^{\gamma }{}_{\gamma }, \\
W = - \frac{3}{8} P^{\gamma }{}_{\gamma }, \\
Z^{\alpha  \beta  \gamma } = \frac{1}{8} \overline{Z}^{\alpha \beta \gamma }, \\
W_{\alpha \beta } = \frac{1}{8} \overline{Z}_{\alpha \beta 9}, \\
Z^{\alpha } = \frac{1}{8} P^{\alpha }{}_{9}. \label{e8trans8}
\end{gather}
Other normalisations can be chosen for the translation generators. However, the above choice of normalisation for the generators is made in order to make contact with the generators found when decomposing the $l_{1}$ generators of $E_{11}$ into GL(3)$\times E_{8}$ \cite{westl1}
\begin{gather*} 
P_{\alpha}, \quad Z^{\alpha \beta }, \quad Z^{\alpha_1 \dots \alpha_5} = \frac{1}{3!} \epsilon^{\alpha_1 \dots \alpha_8} W_{\alpha_6 \dots \alpha_8}, \\
Z^{\alpha_1 \dots \alpha_7, \beta } = \frac{1}{7!} \epsilon^{\alpha_1 \dots \alpha_8} W^{\beta }{}_{\alpha_8}, \quad
Z^{\alpha_1 \dots \alpha_8} = \frac{1}{8!} \epsilon^{\alpha_1 \dots \alpha_8} W, \\
Z^{\alpha_1 \dots \alpha_8, \beta  \gamma  \delta}= \epsilon^{\alpha_1 \dots \alpha_8} Z^{\beta  \gamma \delta}, 
\quad Z^{\alpha_1 \dots \alpha_8, \beta _1 \dots \beta _6} =  \frac{1}{2} \epsilon^{\alpha_1 \dots \alpha_8} \epsilon^{\beta _1 \dots \beta _8} W_{\beta _7 \beta _8}\\
Z^{\alpha_1 \dots \alpha_8, \beta _1 \dots \beta _8, \gamma } = \epsilon^{\alpha_1 \dots \alpha_8} \epsilon^{\beta _1 \dots \beta _8} Z^{\gamma }.
\end{gather*}
The generator $Z^{\alpha_1 \dots \alpha_7, \beta}$ satisfies 
$$Z^{[\alpha_1 \dots \alpha_7, \beta]} = 0$$
because $W^{\alpha}{}_{\beta}$ is traceless. The $l_1$ representation of $E_{11}$ is the highest weight representation where the highest weight corresponds to the $P_1$ translation generator. Recall that the roots of an algebra correspond to the group generators, while the weights of a representation correspond to the translation generators, which generate a particular representation. In \cite{BGPW}, the truncation of $E_{11} \ltimes l_1$ to the SL(5), $E_6$ and $E_7$ motion groups was shown to lead to correct duality-invariant dynamics.    

The commutation relations of generators \eqref{e8trans1}--\eqref{e8trans8} with the $E_{8}$ group generators are found by inserting the SL(8) decomposition of the motion group generators into \eqref{e8mtnalg91}--\eqref{e8mtnalg99}. Here, we list the commutation relations that are required for the non-linear realisation of the $E_8$ motion group. The commutation relations of the GL(8) generator with the translations generators is   
\begin{gather}
[K^{\alpha}{}_{\beta }, P_{\gamma }] = - \delta^{\alpha}_{\gamma } P_{\beta } - \delta^{\alpha}{}_{\beta } P_{\gamma }, \label{comKP} \\
[K^{\alpha}{}_{\beta }, Z^{\gamma \delta}] = 2 \delta^{[\gamma }_{\beta } Z^{|\alpha|\delta]} - \delta^{\alpha}{}_{\beta } Z^{\gamma \delta}, \\ 
[K^{\alpha}{}_{\beta }, W_{\gamma \delta \epsilon}] = - 3 \delta^{\alpha}_{[\gamma } W_{\delta \epsilon]\beta }, \\
[K^{\alpha}{}_{\beta }, W^{\gamma }{}_{\delta}] = \delta^{\gamma }_{\beta } W^{\alpha}{}_{\delta} - \delta^{\alpha}{}_{\delta} W^{\gamma }_{\beta }, \\
[K^{\alpha}{}_{\beta }, W] = 0, \\
[K^{\alpha}{}_{\beta }, Z^{\gamma \delta \epsilon}] = 3 \delta^{[\gamma }_{\beta } Z^{\delta \epsilon]\alpha}, \\
[K^{\alpha}{}_{\beta }, W_{\gamma \delta}] = - 2 \delta^{\alpha}_{[\gamma } W_{|\beta |\delta]} + \delta^{\alpha}{}_{\beta } W_{\gamma \delta}, \\
[K^{\alpha}{}_{\beta }, Z^{\gamma }] = \delta^{\gamma }_{\beta } Z^{\alpha} + \delta^{\alpha}{}_{\beta } Z^{\gamma },
\end{gather}
These commutation relations are needed in order to find the dependence of the generalised metric on the 8-dimensional metric. To find the dependence of the generalised metric on the 3-form and 6-form fields the following commutation relations are required
\begin{gather} 
[R^{\alpha_1 \dots \alpha_3}, P_{\beta }] = 3 \delta^{[\alpha_1}_{\beta } Z^{\alpha_2 \alpha_3]}, \label{crR3uP} \\
[R^{\alpha_1 \dots \alpha_3}, Z^{\beta  \gamma }] = \frac{1}{3!} \epsilon^{\alpha_1 \dots \alpha_3 \beta  \gamma  \delta_1 \dots \delta_3} W_{\delta_1 \dots \delta_3}, \label{crR3uZ2} \\
[R^{\alpha_1 \dots \alpha_3}, W_{\beta _1 \dots \beta _3}] = 18 \delta^{[\alpha_1 \alpha_2}_{[\beta 1 \beta _2} W^{\alpha_3]}{}_{\beta _3]}, \\
[R^{\alpha_1 \dots \alpha_3}, W^{\beta }{}_{\gamma }] = 24 \delta^{[\alpha_1}_{\gamma } Z^{\alpha_2 \alpha_3] \beta } - 3 \delta^{\beta }_{\gamma } Z^{\alpha_1 \dots \alpha_3}, \\
[R^{\alpha_1 \dots \alpha_3}, W] = Z^{\alpha_1 \dots \alpha_3}, \\
[R^{\alpha_1 \dots \alpha_3}, Z^{\beta _1 \dots \beta _3}] = - \frac{1}{2} \epsilon^{\alpha_1 \dots \alpha_3 \beta _1 \dots \beta _3 \gamma  \delta} W_{\gamma \delta}, \\
[R^{\alpha_1 \dots \alpha_3}, W_{\beta  \gamma }] = 6 \delta^{[\alpha_1 \alpha_2}_{\beta  \gamma } Z^{\alpha_3]}, \label{crR3uW2}\\
[R^{\alpha_1 \dots \alpha_6}, P_{\beta }] = \frac{1}{4} \epsilon^{\alpha_1 \dots \alpha_8} W_{\alpha_7 \alpha_8 \beta}, \\
[R^{\alpha_1 \dots \alpha_6}, Z^{\beta  \gamma }] = \epsilon^{\alpha_1 \dots \alpha_6 [\beta | \delta } W^{|\gamma ]}{}_{\delta} - \epsilon^{\alpha_1 \dots \alpha_6 \beta  \gamma  } W, \\
[R^{\alpha_1 \dots \alpha_6}, W_{\beta _1 \dots \beta _3}] = 480 \delta^{[\alpha_1 \dots \alpha_3}_{[\beta _1 \dots \beta _3} Z^{\alpha_4 \dots \alpha_6]}, \\
[R^{\alpha_1 \dots \alpha_6}, W^{\beta }{}_{\gamma }] = 4 \epsilon^{\alpha_1 \dots \alpha_6 \beta  \delta} W_{\gamma  \delta} - \frac{1}{2} \delta^{\beta }_{\gamma } \epsilon^{\alpha_1 \dots \alpha_6 \delta \epsilon} W_{\delta \epsilon}, \\
[R^{\alpha_1 \dots \alpha_6}, W] = - \frac{1}{2} \epsilon^{\alpha_1 \dots \alpha_6 \beta  \gamma } W_{\beta  \gamma }, \\
[R^{\alpha_1 \dots \alpha_6}, Z^{\beta _1 \dots \beta _3}] = \frac{3}{2} \epsilon^{\alpha_1 \dots \alpha_6 [\beta _1 \beta _2 } Z^{\beta _3]}.
\end{gather}
Finally, as the generalised metric is found by conjugating the translation generators by the $E_{8}$ generators corresponding to the positive roots, we also need
\begin{gather} 
[R^{\alpha_1 \dots \alpha_8, \beta }, P_{\gamma }] = - \frac{1}{2} \epsilon^{\alpha_1 \dots \alpha_8} W^{\beta }{}_{\gamma } - \frac{3}{2} \delta^{\beta }_{\gamma } \epsilon^{\alpha_1 \dots \alpha_8} W, \\
[R^{\alpha_1 \dots \alpha_8, \beta }, Z^{\gamma \delta}] = -4 \epsilon^{\alpha_1 \dots \alpha_8} Z^{\beta  \gamma  \delta}, \\
[R^{\alpha_1 \dots \alpha_8, \beta }, W_{\gamma _1 \dots \gamma _3}] = 12 \epsilon^{\alpha_1 \dots \alpha_8} \delta^{\beta }_{[\gamma _1} W_{\gamma _2 \gamma _3]}, \\
[R^{\alpha_1 \dots \alpha_8, \beta }, W^{\gamma }{}_{\delta}] = 4 \delta^{\beta }_{\delta} \epsilon^{\alpha_1 \dots \alpha_8} Z^{\gamma } - \frac{1}{2} \delta^{\gamma }_{\delta} \epsilon^{\alpha_1 \dots \alpha_8} Z^{\beta }, \\
[R^{\alpha_1 \dots \alpha_8, \beta }, W] = \frac{3}{2} \epsilon^{\alpha_1 \dots \alpha_8} Z^{\beta }. \label{comR81W}
\end{gather}
Some of the commutation relations involving the generators corresponding to negative roots are listed below:
\begin{gather} 
[R_{\alpha_1 \dots \alpha_3}, Z^{\beta  \gamma}] = 6 \delta^{\beta \gamma}_{[\alpha_1 \alpha_2} P_{\alpha_3]}, \label{crR3dZ2}\\
[R_{\alpha_1 \dots \alpha_3}, W_{\beta _1 \dots \beta _3}] = \frac{1}{2} \epsilon_{\alpha_1 \dots \alpha_3 \beta _1 \dots \beta _3 \gamma \delta} Z^{\gamma \delta}, \label{crR3dW3} \\
[R_{\alpha_1 \dots \alpha_3}, W^{\beta }{}_{\gamma }] = 3 \delta_{[\alpha_1}^{\beta} W_{\alpha_2 \alpha_3] \gamma } - \frac{3}{8} \delta^{\beta }_{\gamma } W_{\alpha_1 \dots \alpha_3}, \\
[R_{\alpha_1 \dots \alpha_3}, W] = \frac{1}{8} W_{\alpha_1 \dots \alpha_3}, \\
[R_{\alpha_1 \dots \alpha_3}, Z^{\beta _1 \dots \beta _3}] = \frac{9}{4} \delta^{[\beta_1 \beta_2}_{[\alpha_1 \alpha_2} W^{\beta_3]}{}_{\alpha_3]} +\frac{3}{4} \delta^{\beta_1 \dots \beta_3}_{\alpha_1 \dots \alpha_3} W, \\
[R^{\alpha_1 \dots \alpha_3}, W_{\beta  \gamma }] = - \frac{1}{3!} \epsilon_{\alpha_1 \dots \alpha_3 \beta \gamma \delta_1 \dots \delta_3} Z^{\delta_1 \dots \delta_3}, \\
[R_{\alpha_1 \dots \alpha_3}, Z^{\beta }] = 3 \delta_{[\alpha_1}^{\beta} W_{\alpha_2 \alpha_3] }. \label{crR3dZ}
\end{gather}
We take the translation generators to mutually-commute.
 
The normalisations of these generators are needed in the calculation of the generalised metric using non-linear realisation. Denoting the Cartan involution of a generator $X$ by $X^*,$ 
we can define an inner product on the representation space generated by the translation generators \cite{BGPW}
$$(A, B^*) \in \mathbb{R},$$
where $A$ and $B$ are translation generators. The inner product is $E_8$ invariant 
\begin{equation}
 ([X,A], B^*) = - (A, [X, B^*]),
 \label{inninv}
\end{equation}
where $X$ is an $E_8$ generator.

The Cartan involution interchanges negative and positive roots. Therefore,
$$ K^{*}{}^{\, \alpha}{}_{\beta} \sim K^{\beta}{}_{\alpha}, \;\;\; R^{*}{\,}^{\alpha_1 \dots \alpha_3} \sim R_{\alpha_1 \dots \alpha_3}, \;\;\; 
R^{*}{\,}^{\alpha_1 \dots \alpha_6} \sim R_{\alpha_1 \dots \alpha_6}, \;\;\; R^{*}{\,}^{\alpha_1 \dots \alpha_8, \beta} \sim R_{\alpha_1 \dots \alpha_8, \beta}.$$
We define 
\begin{equation}
 R^{*}{\,}^{\alpha_1 \dots \alpha_3} = - R_{\alpha_1 \dots \alpha_3}. 
\label{CartaninvR3}
\end{equation}
The relative signs of the Cartan involution of the other generators is fixed by the above relation and consistency with the $E_8$ algebra. For example, the Cartan involution of equation \eqref{crR3uR3d}
\begin{align*}
 [ R^{*}{\,}^{\alpha_1 \dots \alpha_3}, R^{*}{\,}_{\beta_1 \dots \beta_3}] &=  18 \delta^{[\alpha_1 \alpha_2}{}_{[\beta _1 \beta _2} K^{*}{\,}^{\alpha_3]}{}_{\beta 3]} - 2 \delta^{\alpha_1 \dots \alpha_3}_{\beta _1 \dots \beta _3} K^{*}{\,}, \\
 \implies \qquad  
 [ R_{\alpha_1 \dots \alpha_3}, R^{\beta_1 \dots \beta_3}] &=  18 \delta^{[\alpha_1 \alpha_2}{}_{[\beta _1 \beta _2} K^{*}{\,}^{\alpha_3]}{}_{\beta 3]} - 2 \delta^{\alpha_1 \dots \alpha_3}_{\beta _1 \dots \beta _3} K^{*}{\,}, \\
  \implies \qquad  
 - [R^{\beta_1 \dots \beta_3}, R_{\alpha_1 \dots \alpha_3}] &=  18 \delta^{[\alpha_1 \alpha_2}{}_{[\beta _1 \beta _2} K^{*}{\,}^{\alpha_3]}{}_{\beta 3]} - 2 \delta^{\alpha_1 \dots \alpha_3}_{\beta _1 \dots \beta _3} K^{*}{\,},
\end{align*}
which from equation \eqref{crR3uR3d} gives that 
\begin{equation}
K^{*}{}^{\, \alpha}{}_{\beta} = - K^{\beta}{}_{\alpha}. 
\label{CartaninvK}
\end{equation}
Similarly, the Cartan involution of the rest of the generators are 
\begin{equation}
 R^{*}{\,}^{\alpha_1 \dots \alpha_6} = R_{\alpha_1 \dots \alpha_6},
\label{CartaninvR6}
\end{equation}
\begin{equation}
R^{*}{\,}^{\alpha_1 \dots \alpha_8, \beta} = - R_{\alpha_1 \dots \alpha_8, \beta}. 
\label{CartaninvR81}
\end{equation}

Now to find the normalisation of the generators, we first define
\begin{equation}
  (P_{\alpha}, P^{*}{}^{\beta})= \delta^{\beta}_{\alpha}.
\end{equation}
The normalisation of all other translation generators are now fixed. For example, consider 
\begin{align*}
 (Z^{\alpha \beta}, Z^{*}_{\gamma \delta}) &= \frac{1}{2} (Z^{\alpha \beta}, Z^{*}_{[\gamma \delta}) \delta^{\eta}_{\eta]}, \\
 &= \frac{1}{2} (Z^{\alpha \beta}, - \frac{1}{3}[R_{\gamma \delta \eta}, P^{*}{\,}^{\eta}])
\end{align*}
from the Cartan involution of equation \eqref{crR3uP}. The $E_8$ invariance of the inner product, \eqref{inninv}, allows us to write the inner product in terms of equation \eqref{crR3dZ2}, hence giving
$$ (Z^{\alpha \beta}, Z^{*}{}_{\gamma \delta})= 2 \delta^{\alpha \beta}_{\gamma \delta}.$$
Similarly, from equations \eqref{crR3uZ2}--\eqref{crR3uW2} and equations \eqref{crR3dW3}--\eqref{crR3dZ},
the inner product of the rest of the translation generators is   
\begin{gather}
 (W_{\alpha_1 \dots \alpha_3}, W^{*}{}^{\beta_1 \dots \beta_3})= 6 \delta^{\beta_1 \dots \beta_3}_{\alpha_1 \dots \alpha_3}, \\
(W^{\alpha}{}_{\beta}, W^{*}{}_{\gamma}{}^{\delta}) = \delta^{\alpha}_{\gamma} \delta^{\delta}_{\beta} - \frac{1}{8} \delta^{\alpha}_{\beta} \delta^{\delta}_{\gamma}, \\
(W, W^{*}) = \frac{1}{8}, \\
(Z^{\alpha_1 \dots \alpha_3}, Z^{*}{}_{\beta_1 \dots \beta_3}) = \frac{3}{32} \delta^{\alpha_1 \dots \alpha_3}_{\beta_1 \dots \beta_3}, \\
(W_{\alpha \beta}, W^{*}{}^{\gamma \delta})= \frac{1}{32} \delta^{\gamma \delta}_{\alpha \beta}, \\
(Z^{\alpha}, Z^{*}{}_{\beta})= \frac{1}{64} \delta^{\alpha}_{\beta}.
\end{gather}

\newpage

\section{The Generalised Vielbein}
\label{appviel}

In this appendix we give the components of $L^{A}{}_{B},$ which is related to the generalised vielbein by equation \eqref{genvielrel}.
The components of $L^{A}{}_{B},$ see figure \ref{tableviel}, are
\begin{align}
 (L11)^{a}{}_{\; b} &=    \delta^{a}_{b}, \\
 (L21)_{d_1 d_2}{}_{\; b} &= -\frac{1}{\sqrt{2}} C_{d_1 d_2 b}, \\
 (L31)^{g_1 g_2 g_3}{}_{\; b}& = - \frac{\sqrt{3}}{2 \sqrt{2}} \delta^{[g_1}_{b} U^{g_2 g_3]} - \frac{1}{4 \sqrt{6}} X^{g_1 \dots g_3}{}_{b}, \\
 (L41)_{j_1}{}^{j_2}{}_{\; b} &= \frac{1}{24} X^{u_1 u_2 j_2}{}_{j_1} C_{u_1 u_2 b} + \frac{1}{2} C_{u j_1 b} U^{u j_2} - \frac{1}{16} \delta^{j_2}_{j_1} C_{u_1 u_2 b} U^{u_1 u_2} + \frac{1}{2} \delta^{j_2}_{b} Y_{j_1} - \frac{1}{16} \delta^{j_2}_{j_1} Y_{b}, \\
 (L51)_{\; b} &= \frac{3}{4 \sqrt{2}} Y_{b} - \frac{1}{4 \sqrt{2}} C_{u_1 u_2 b} U^{u_1 u_2}, \\
 (L61)_{m_1 m_2 m_3}{}_{\; b} &= -\frac{\sqrt{3}}{2 \sqrt{2}} C_{b [m_1 m_2} Y_{m_3]} + \frac{1}{16 \sqrt{6}} C_{u_1 u_2 b m_1 m_2 m_3} U^{u_1 u_2} \notag \\ 
& \quad \;\; + \frac{1}{48 \sqrt{6}} X^{u_1 u_2 u_3}{}_{b} C_{u_1 u_2 u_3 m_1 m_2 m_3} - \frac{1}{32 \sqrt{6}} C_{u_1 [m_1 m_2} C_{m_3] u_2 u_3} X^{u_1 u_2 u_3}{}_{b}, \\
 (L71)^{q_1 q_2}{}_{\; b} &= \frac{3}{4 \sqrt{2}} \delta^{[u}_{b} U^{q_1 q_2]} Y_{u} + \frac{1}{8 \sqrt{2}} X^{q_1 q_2 u}{}_{b} Y_{u}  -\frac{1}{4 \sqrt{2}} C_{u_1 u_2 b} U^{u_1 q_1} U^{u_2 q_2} \notag \\
&\quad \;\;  + \frac{1}{24 \sqrt{2}} C_{u_1 u_2 u_3} X^{u_1 u_2 [q_1}{}_{b} U^{q_2] u_3} + \frac{1}{960 \sqrt{2}} C_{u_1 u_2 u_3} X^{u_1 q_1 q_2}{}_{t} X^{t u_2 u_3}{}_{b}, \\
(L81)_{x}{}_{\; b} &= -\frac{1}{4} Y_{x} Y_{b} - \frac{1}{4} C_{u_1 x b} U^{u_1 u_2} Y_{u_2} + \frac{1}{8} C_{u_1 u_2 b} U^{u_1 u_2} Y_{x} - \frac{1}{48} C_{x u_1 u_2} X^{u_1 u_2 u_3}{}_{b} Y_{u_3} \notag \\
& \quad \;\; + \frac{1}{192} C_{xbu_1 \dots u_4} U^{u_1 u_2} U^{u_3 u_4} - \frac{1}{384} X^{u_1 \dots u_3}{}_{b} U^{u_4 u_5} C_{u_1 \dots u_5 x} \notag \\
& \quad \;\; + \frac{1}{128} C_{u_1 [t_1 t_2} C_{x] u_2 u_3} X^{u_1 \dots u_3}{}_{b} U^{t_1 t_2} - \frac{1}{16 (6!)} C_{u_1 u_2 t_1} C_{u_3 t_2 t_3} X^{u_1 \dots u_3}{}_{x} X^{t_1 \dots t_3}{}_{b}, \\
(L22)_{d_1 d_2}{}^{\; e_1 e_2} &= \delta^{e_1 e_2}_{d_1 d_2}, \\
(L32)^{g_1 g_2 g_3}{}^{\; e_1 e_2} &= \frac{1}{2 \sqrt{3}} V^{g_1 g_2 g_3 e_1 e_2}, \\
(L42)_{j_1}{}^{j_2}{}^{\; e_1 e_2} &= - \frac{1}{4 \sqrt{2}} X^{j_2 e_1 e_2}{}_{j_1} + \frac{1}{\sqrt{2}} U^{j_2 [ e_1} \delta^{e_2]}_{j_1} + \frac{1}{8 \sqrt{2}} \delta^{j_2}_{j_1} U^{e_1 e_2}, \\
(L52)^{\; e_1 e_2} &= \frac{1}{4} U^{e_1 e_2},
\end{align}
\begin{align}
(L62)_{m_1 m_2 m_3}{}^{\; e_1 e_2} &= \frac{\sqrt{3}}{2} Y_{[m_1} \delta^{e_1 e_2}_{m_2 m_3]} - \frac{1}{24 \sqrt{3}} V^{u_1 u_2 u_3 e_1 e_2}  C_{u_1 u_2 u_3 m_1 m_2 m_3} \notag \\
& \qquad \qquad \qquad \qquad\qquad\qquad \qquad \qquad + \frac{1}{8 \sqrt{3}} C_{u [m_1 m_2} X^{u e_1 e_2}{}_{m_3]}, \\
(L72)^{q_1 q_2}{}^{\; e_1 e_2} &= - \frac{1}{4} V^{q_1 q_2 e_1 e_2 u} Y_{u} + \frac{1}{4} U^{[q_1 | e_1} U^{| q_2] e_2} - \frac{1}{8} X^{e_1 e_2 [ q_1}{}_{u} U^{q_2] u} - \frac{1}{192} X^{q_1 q_2 u}{}_{t} X^{e_1 e_2 t}{}_{u}, \\
(L82)_{x}{}^{\; e_1 e_2} &= - \frac{1}{2 \sqrt{2}} U^{u [e_1} Y_{u} \delta^{e_2]}_{x} + \frac{1}{8 \sqrt{2}} X^{e_1 e_2 u}{}_{x} Y_{u} - \frac{1}{4 \sqrt{2}} U^{e_1 e_2} Y_{x}  \notag \\
& \quad \;\; + \frac{1}{96 \sqrt{2}} V^{e_1 e_2 u_1 u_2 u_3} U^{ u_4 u_5} C_{u_1 \dots u_5 x} - \frac{1}{16 \sqrt{2}} C_{t [ u_1 u_2} X^{ t e_1 e_2}{}_{x]} U^{u_1 u_2} \notag \\
& \quad \;\;  + \frac{1}{960 \sqrt{2}} C_{t u_1 u_2} X^{u_1 u_2 u_3}{}_{x} X^{e_1 e_2 t}{}_{u_3}
\end{align} 
\begin{align}
(L33)^{g_1 g_2 g_3}{}_{\; h_1 h_2 h_3} &= g^{-1/2} \delta^{g_1 g_2 g_3}_{h_1 h_2 h_3},  \\
(L43)_{j_1}{}^{\; j_2}{}_{\; h_1 h_2 h_3} &= - \sqrt{\frac{3}{2}} g^{-1/2} \left( C_{j_1 [h_1 h_2 } \delta^{j_2}_{h_{3}]} - \frac{1}{8} \delta^{j_2}_{j_1} C_{h_1 h_2 h_3}  \right), \\
(L53)_{\; h_1 h_2 h_3} &=  - \frac{1}{4 \sqrt{3}} g^{-1/2} C_{h_1 h_2 h_3}, \\ 
(L63)_{m_1 m_2 m_3}{}_{\; h_1 h_2 h_3} &= \frac{1}{12} g^{-1/2} \left(  C_{m_1 m_2 m_3 h_1 h_2 h_3} - C_{m_1 m_2 m_3} C_{h_1 h_2 h_3} + 9 C_{[m_1 m_2 | [| h_1 } C_{h_2 h_3|] |m_3]}  \right), \\
(L73)^{q_1 q_2}{}_{\; h_1 h_2 h_3} &= - \frac{\sqrt{3}}{2} g^{-1/2} \left( Y_{[h_1} \delta^{q_1 q_2}_{h_2 h_3]} + C_{u [h_1 h_2 } U^{u [q_1} \delta^{q_2]}_{h_3]} + \frac{1}{6} C_{h_1 h_2 h_3} U^{q_1 q_2} \right. \notag \\
& \left. \qquad\qquad\qquad \;\;\;  + \frac{1}{12} X^{u q_1 q_2}{}_{[h_1} C_{h_2 h_3] u} \right),     \\
(L83)_{x}{}_{\; h_1 h_2 h_3} &= \frac{\sqrt{3}}{2 \sqrt{2}} g^{-1/2} \left( C_{x [h_1 h_2} Y_{h_3]} + \frac{1}{24} U^{u_1 u_2} C_{h_1 h_2 h_3 x u_1 u_2} + \frac{1}{12} C_{h_1 h_2 h_3} C_{x u_1 u_2} U^{u_1 u_2} \right. \notag \\
&  \qquad \qquad\qquad \;\;\; - \frac{1}{4} C_{u_1 u_2 [h_1} C_{h_2 h_3] x} U^{u_1 u_2} - \frac{1}{2} C_{x u_1 [h_1} C_{h_2 h_3] u_2} U^{u_1 u_2} \notag \\
& \left. \qquad \qquad\qquad \;\;\; + \frac{1}{48} X^{u_1 u_2 u_3}{}_{x} C_{u_1 u_2 [h_1} C_{h_2 h_3] u_3} \right),
\end{align} 
\begin{align}
(L44)_{j_1}{}^{j_2}{}^{\; k_1}{}_{k_2} &= g^{-1/2} \left( \delta^{j_2}_{k_2} \delta^{k_1}_{j_1} - \frac{1}{8} \delta^{j_2}_{j_1} \delta^{k_1}_{k_2} \right), \\
(L54)^{\; k_1}{}_{k_2} &= 0, 
\end{align} 
\begin{align}
(L64)_{m_1 m_2 m_3}{}^{\; k_1}{}_{k_2} &= - \sqrt{\frac{3}{2}} g^{-1/2} \left( C_{k_2 [m_1 m_2} \delta^{k_1}_{m_3]} - \frac{1}{8} \delta^{k_1}_{k_2} C_{m_1 m_2 m_3} \right), \\
(L74)^{q_1 q_2}{}^{\; k_1}{}_{k_2} &=  \frac{1}{\sqrt{2}} g^{-1/2} \left(U^{k_1 [q_1} \delta^{q_2]}_{k_2} + \frac{1}{8} \delta^{k_1}_{k_2} U^{q_1 q_2} + \frac{1}{4} X^{k_1 q_1 q_2}{}_{k_2}  \right), \\
(L84)_{x}{}^{\; k_1}{}_{k_2} &= - \frac{1}{2} g^{-1/2} \left( Y_{k_2} \delta^{k_1}_{x} - \frac{1}{8} \delta^{k_1}_{k_2} Y_{x} - \frac{3}{2} \delta^{k_1}_{[x} C_{u_1 u_2] k_2} U^{u_1 u_2} + \frac{3}{16} \delta^{k_1}_{k_2} C_{x u_1 u_2} U^{u_1 u_2} \right. \notag \\
& \left.  \qquad \qquad\qquad \;\;\; + \frac{1}{12} X^{k_1 u_1 u_2}{}_{k_2} C_{x u_1 u_2} \right), \\
L55 &= g^{-1/2},  \\ 
(L65)_{m_1 m_2 m_3} &= - \frac{1}{4 \sqrt{3}} g^{-1/2} C_{m_1 m_2 m_3}, \\
(L75)^{q_1 q_2} &=  \frac{1}{4} g^{-1/2} U^{q_1 q_2}, \\
(L85)_{x} &= - \frac{3}{4 \sqrt{2}} g^{-1/2} \left( Y_{x} - \frac{1}{6} C_{x u_1 u_2} U^{u_1 u_2} \right),
\end{align} 
\begin{align}
(L66)_{m_1 m_2 m_3}{}^{\; n_1 n_2 n_3} &= g^{-1/2} \delta^{n_1 n_2 n_3}_{m_1 m_2 m_3}, \\
(L76)^{q_1 q_2}{}^{\; n_1 n_2 n_3} &=  - \frac{1}{2 \sqrt{3}} g^{-1/2} V^{q_1 q_2 n_1 n_2 n_3}, \\
(L86)_{x}{}^{\; n_1 n_2 n_3} &= - \frac{\sqrt{3}}{2 \sqrt{2}} g^{-1/2} \left( U^{[n_1 n_2} \delta^{n_3]}_{x} - \frac{1}{6} X^{n_1 n_2 n_3}{}_{x} \right),
\end{align} 
\begin{align}
(L77)^{q_1 q_2}{}_{\; r_1 r_2} &=  g^{-1} \delta^{q_1 q_2}_{r_1 r_2}, \\
(L87)_{x}{}_{\; r_1 r_2} &= - \frac{1}{\sqrt{2}} g^{-1} C_{x r_1 r_2}, \\
(L88)_{x}{}^{\; y} &=  g^{-1} \delta^{y}_{x}.
\end{align} 
All of the lowercase Latin letters denote SL(8) indices. In the above expressions $g$ is the determinant of the spatial metric, 
\begin{equation}
 V^{a_1 \dots a_5}= \frac{1}{3!} \epsilon^{a_1 \dots a_8} C_{a_6 \dots a_8},
\label{Vdef}
\end{equation}
\begin{equation}
  X^{a_1 \dots a_3}{}_{b} = V^{a_1 \dots a_5} C_{a_4 a_5 b},
\label{Xdef}
\end{equation}
\begin{equation} W^{a_1 a_2} = \frac{1}{6!} \epsilon^{a_1 \dots a_8} C_{a_3 \dots a_8},\end{equation}
\begin{equation} Y_{b} = \frac{1}{8!} \epsilon^{a_1 \dots a_8} C_{a_1 \dots a_8, b}.\end{equation}
The $\epsilon$ tensor is the alternating tensor in eight dimensions.

\begin{sidewaystable}
\centering
\begin{tabular}{l}
$
\begin{pmatrix}
 (L11)^{a}{}_{\; b} & 0 & 0 & 0 & 0 & 0 & 0 & 0 \\
&&&&&&&\\
&&&&&&&\\
 (L21)_{d_1 d_2}{}_{\; b} & (L22)_{d_1 d_2}{}^{\; e_1 e_2} & 0 & 0 & 0 & 0 & 0 & 0 \\
&&&&&&&\\
&&&&&&&\\
 (L31)^{g_1 \dots g_3}{}_{\; b} & (L32)^{g_1 \dots g_3}{}^{\; e_1 e_2} & (L33)^{g_1 \dots g_3}{}_{\; h_1 \dots h_3} & 0 & 0 & 0 & 0 & 0 \\
&&&&&&&\\
&&&&&&&\\
 (L41)_{j_1}{}^{j_2}{}_{\; b} & (L42)_{j_1}{}^{j_2}{}^{\; e_1 e_2} & (L43)_{j_1}{}^{j_2}{}_{\; h_1 \dots h_3} & (L44)_{j_1}{}^{j_2}{}^{\; k_1}{}_{k_2} & 0 & 0 & 0 & 0 \\
&&&&&&&\\
&&&&&&&\\
 (L51)_{\; b} & (L52)^{\; e_1 e_2} & (L53)_{\; h_1 \dots h_3} & (L54)^{\;k_1}{}_{k_2} & L55 & 0 & 0 & 0 \\
&&&&&&&\\
&&&&&&&\\
 (L61)_{m_1 \dots m_3}{}_{\; b} & (L62)_{m_1 \dots m_3}{}^{\; e_1 e_2} & (L63)_{m_1 \dots m_3}{}_{\; h_1 \dots h_3} & (L64)_{m_1 \dots m_3}{}^{\; k_1}{}_{k_2} & (L65)_{m_1 \dots m_3} & (L66)_{m_1 \dots m_3}{}^{\; n_1 \dots n_3} & 0 & 0 \\
&&&&&&&\\
&&&&&&&\\
(L71)^{q_1 q_2}{}_{\; b} & (L72)^{q_1 q_2}{}^{\; e_1 e_2} & (L73)^{q_1 q_2}{}_{\; h_1 \dots h_3} & (L74)^{q_1 q_2}{}^{\; k_1}{}_{k_2} & (L75)^{q_1 q_2} & (L76)^{q_1 q_2}{}^{\; n_1 \dots n_3} & (L77)^{q_1 q_2}{}_{\; r_1 r_2} & 0 \\
&&&&&&&\\
&&&&&&&\\
 (L81)_{x}{}_{\; b} & (L82)_{x}{}^{\; e_1 e_2} & (L83)_{x}{}_{\; h_1 \dots h_3} & (L84)_{x}{}^{\; k_1}{}_{k_2} & (L85)_{x} & (L86)_{x}{}^{\; n_1 \dots n_3} & (L87)_{x}{}_{\; r_1 r_2} & (L88)_{x}{}^{\; y}
\end{pmatrix}
$ \vspace{6mm}
\end{tabular} 
\captionsetup{tablename=Figure}
\caption{The generalised vielbein}
\label{tableviel} \end{sidewaystable} 

Similarly, the components of the inverse generalised vielbein $E^{A}{}_{B},$ see figure \ref{tableinvviel}, are
\begin{align}
 (E11)^{a}{}_{\; b} &=    \delta^{a}_{b}, \\
 (E21)_{d_1 d_2}{}_{\; b} &= \frac{1}{\sqrt{2}} C_{d_1 d_2 b}, \\
 (E31)^{g_1 g_2 g_3}{}_{\; b}& = \frac{\sqrt{3}}{2 \sqrt{2}} g^{1/2} \left(  \delta^{[g_1}_{b} U^{g_2 g_3]} - \frac{1}{6} X^{g_1 \dots g_3}{}_{b} \right), \\
 (E41)_{j_1}{}^{j_2}{}_{\; b} &= - \frac{1}{24} g^{1/2} \left( X^{u_1 u_2 j_2}{}_{j_1} C_{u_1 u_2 b} -12 C_{u j_1 b} U^{u j_2} - 6 \delta^{j_2}_{b} C_{u_1 u_2 j_1} U^{u_1 u_2} \right. \notag \\
  & \qquad \qquad \qquad \; \;\; \left. + \frac{9}{4} \delta^{j_2}_{j_1} C_{u_1 u_2 b} U^{u_1 u_2} + 12 \delta^{j_2}_{b} Y_{j_1} - \frac{3}{2} \delta^{j_2}_{j_1} Y_{b} \right), \\
 (E51)_{\; b} &= - \frac{3}{4 \sqrt{2}} g^{1/2} \left(  Y_{b} - \frac{1}{6} C_{u_1 u_2 b} U^{u_1 u_2} \right), \\
 (E61)_{m_1 m_2 m_3}{}_{\; b} &=  -\frac{\sqrt{3}}{2 \sqrt{2}} g^{1/2} \left(  C_{b [m_1 m_2} Y_{m_3]} - \frac{1}{24} C_{u_1 u_2 b m_1 m_2 m_3} U^{u_1 u_2}  \right. \notag \\ 
& \qquad \qquad \; \qquad \;\; + \frac{5}{6} C_{[b u_1 u_2} C_{m_1 m_2 m_3]} U^{u_1 u_2} \notag \\
& \qquad \qquad \qquad \;\;\; \left. + \frac{1}{48} C_{u_1 [m_1 m_2} C_{m_3] u_2 u_3} X^{u_1 u_2 u_3}{}_{b} \right), \\
 (E71)^{q_1 q_2}{}_{\; b} &= \frac{3}{4 \sqrt{2}} g \left(\delta^{[u}_{b} U^{q_1 q_2]} Y_{u} - \frac{1}{6} X^{q_1 q_2 u}{}_{b} Y_{u}  + \frac{1}{72} V^{q_1 q_2 u_1 u_2 u_3 }C_{b u_1 \dots u_5} U^{u_4 u_5} \right. \notag \\
&\qquad \qquad \; \;\;  \left. + \frac{1}{12} C_{u_1 [u_2 u_3} X^{u_1 q_1 q_2}{}_{b]} U^{u_2 u_3} - \frac{1}{720} C_{u_1 u_2 u_3} X^{u_1 q_1 q_2}{}_{t} X^{t u_2 u_3}{}_{b} \right), \\
(E81)_{x}{}_{\; b} &=  - \frac{1}{4} g \left( Y_{x} Y_{b} - C_{u_1 x b} U^{u_1 u_2} Y_{u_2} - \frac{1}{2} C_{x u_1 u_2} U^{u_1 u_2} Y_{b}  \right. \notag \\
& \qquad \; \;\; + \frac{1}{12} C_{x u_1 u_2} X^{u_1 u_2 u_3}{}_{b} Y_{u_3} + \frac{1}{48} C_{xbu_1 \dots u_4} U^{u_1 u_2} U^{u_3 u_4}  \notag \\
& \qquad \;\;\; + \frac{1}{96} X^{u_1 \dots u_3}{}_{x} U^{u_4 u_5} C_{u_1 \dots u_5 b} - \frac{1}{32} C_{u_1 [t_1 t_2} C_{b] u_2 u_3} X^{u_1 \dots u_3}{}_{x} U^{t_1 t_2} \notag \\
& \qquad \;\;\; \left.+ \frac{1}{4 (6!)} C_{u_1 u_2 t_1} C_{u_3 t_2 t_3} X^{u_1 \dots u_3}{}_{x} X^{t_1 \dots t_3}{}_{b}\right) ,
\end{align}
\begin{align}
(E22)_{d_1 d_2}{}^{\; e_1 e_2} &= \delta^{e_1 e_2}_{d_1 d_2}, \\
(E32)^{g_1 g_2 g_3}{}^{\; e_1 e_2} &= - \frac{1}{2 \sqrt{3}} g^{1/2} V^{g_1 g_2 g_3 e_1 e_2}, \\
(E42)_{j_1}{}^{j_2}{}^{\; e_1 e_2} &= - \frac{1}{4 \sqrt{2}} g^{1/2} \left( X^{j_2 e_1 e_2}{}_{j_1} + 4 U^{j_2 [ e_1} \delta^{e_2]}_{j_1} + \frac{1}{2} \delta^{j_2}_{j_1} U^{e_1 e_2} \right), \\
(E52)^{\; e_1 e_2} &= - \frac{1}{4} g^{1/2} U^{e_1 e_2}, \\ 
(E62)_{m_1 m_2 m_3}{}^{\; e_1 e_2} &= - \frac{\sqrt{3}}{2} g^{1/2} \left( Y_{[m_1} \delta^{e_1 e_2}_{m_2 m_3]} + C_{u [m_1 m_2} U^{u [e_1} \delta^{e_2]}_{m_3]} \right. \notag \\
& \qquad \qquad \qquad \; \;\;\; \left. + \frac{1}{6} C_{m_1 m_2 m_3} U^{e_1 e_2} + \frac{1}{12} C_{u [m_1 m_2} X^{u e_1 e_2}{}_{m_3]} \right), \\
(E72)^{q_1 q_2}{}^{\; e_1 e_2} &= - \frac{1}{4} g \left( V^{q_1 q_2 e_1 e_2 u} Y_{u} - U^{[q_1 | e_1} U^{| q_2] e_2} \right. \notag \\
&\qquad \qquad \;\;\;\; \left. + \frac{1}{2} X^{q_1 q_2 [ e_1}{}_{u} U^{e_2] u} + \frac{1}{48} X^{q_1 q_2 u}{}_{t} X^{e_1 e_2 t}{}_{u}\right), \\
(E82)_{x}{}^{\; e_1 e_2} &= - \frac{1}{2 \sqrt{2}} g \left( U^{u [e_1} Y_{u} \delta^{e_2]}_{x} + \frac{1}{4} X^{e_1 e_2 u}{}_{x} Y_{u} + \frac{1}{2} U^{e_1 e_2} Y_{x} \right.  \notag \\
& \qquad \qquad \quad \;\;\;\; - \frac{1}{2} C_{u_1 u_2 x} U^{e_1 u_1} U^{e_2 u_2} + \frac{1}{12} C_{u_1 u_2 u_3} X^{u_1 u_2 [e_1 }{}_{x} U^{e_2] u_3} \notag \\
& \qquad \qquad \quad \;\;\;\;  \left. + \frac{1}{480} C_{t u_1 u_2} X^{u_1 u_2 u_3}{}_{x} X^{e_1 e_2 t}{}_{u_3} \right),
\end{align} 
\begin{align}
(E33)^{g_1 g_2 g_3}{}_{\; h_1 h_2 h_3} &= g^{1/2} \delta^{g_1 g_2 g_3}_{h_1 h_2 h_3},  \\
(E43)_{j_1}{}^{\; j_2}{}_{\; h_1 h_2 h_3} &= \sqrt{\frac{3}{2}} g^{1/2} \left( C_{j_1 [h_1 h_2 } \delta^{j_2}_{h_{3}]} - \frac{1}{8} \delta^{j_2}_{j_1} C_{h_1 h_2 h_3}  \right), \\
(E53)_{\; h_1 h_2 h_3} &=  \frac{1}{4 \sqrt{3}} g^{1/2} C_{h_1 h_2 h_3}, \\ 
(E63)_{m_1 m_2 m_3}{}_{\; h_1 h_2 h_3} &= - \frac{1}{12} g^{1/2} \left(  C_{m_1 m_2 m_3 h_1 h_2 h_3} + C_{m_1 m_2 m_3} C_{h_1 h_2 h_3} - 9 C_{[m_1 m_2 | [| h_1 } C_{h_2 h_3|] |m_3]}  \right), \\
(E73)^{q_1 q_2}{}_{\; h_1 h_2 h_3} &= \frac{\sqrt{3}}{2} g \left( Y_{[h_1} \delta^{q_1 q_2}_{h_2 h_3]} - \frac{1}{36} V^{q_1 q_2 u_1 u_2 u_3} C_{u_1 u_2 u_3 h_1 h_2 h_3}  \right. \notag \\
& \left. \qquad\qquad \;\;\;  + \frac{1}{12} X^{u q_1 q_2}{}_{[h_1} C_{h_2 h_3] u} \right),     \\
(E83)_{x}{}_{\; h_1 h_2 h_3} &= \frac{\sqrt{3}}{2 \sqrt{2}} g \left( C_{x [h_1 h_2} Y_{h_3]} + \frac{1}{24} U^{u_1 u_2} C_{h_1 h_2 h_3 x u_1 u_2} - \frac{1}{72} X^{u_1 u_2 u_3}{}_{x} C_{u_1 u_2 u_3 h_1 h_2 h_3} \right. \notag \\
& \left. \qquad\qquad \qquad \;\;\; + \frac{1}{48} X^{u_1 u_2 u_3}{}_{x} C_{u_1 u_2 [h_1} C_{h_2 h_3] u_3} \right),
\end{align} 
\begin{align}
(E44)_{j_1}{}^{j_2}{}^{\; k_1}{}_{k_2} &= g^{1/2} \left( \delta^{j_2}_{k_2} \delta^{k_1}_{j_1} - \frac{1}{8} \delta^{j_2}_{j_1} \delta^{k_1}_{k_2} \right), \\
(E54)^{\; k_1}{}_{k_2} &= 0,  \\ 
(E64)_{m_1 m_2 m_3}{}^{\; k_1}{}_{k_2} &= \sqrt{\frac{3}{2}} g^{1/2} \left( C_{k_2 [m_1 m_2} \delta^{k_1}_{m_3]} - \frac{1}{8} \delta^{k_1}_{k_2} C_{m_1 m_2 m_3} \right), \\
(E74)^{q_1 q_2}{}^{\; k_1}{}_{k_2} &= - \frac{1}{\sqrt{2}} g \left(U^{k_1 [q_1} \delta^{q_2]}_{k_2} + \frac{1}{8} \delta^{k_1}_{k_2} U^{q_1 q_2} - \frac{1}{4} X^{k_1 q_1 q_2}{}_{k_2}  \right), \\
(E84)_{x}{}^{\; k_1}{}_{k_2} &= \frac{1}{2} g \left( Y_{k_2} \delta^{k_1}_{x} - \frac{1}{8} \delta^{k_1}_{k_2} Y_{x} + C_{x u k_2} U^{u k_1} - \frac{1}{8} \delta^{k_1}_{k_2} C_{x u_1 u_2} U^{u_1 u_2} \right. \notag \\
& \left.  \qquad \qquad\qquad \;\;\; + \frac{1}{12} X^{k_1 u_1 u_2}{}_{k_2} C_{x u_1 u_2} \right),
\end{align} 
\begin{align}
E55 &= g^{1/2},  \\ 
(E65)_{m_1 m_2 m_3} &= \frac{1}{4 \sqrt{3}} g^{1/2} C_{m_1 m_2 m_3}, \\
(E75)^{q_1 q_2} &= - \frac{1}{4} g U^{q_1 q_2}, \\
(E85)_{x} &= \frac{3}{4 \sqrt{2}} g \left( Y_{x} - \frac{1}{3} C_{x u_1 u_2} U^{u_1 u_2} \right),
\end{align} 
\begin{align}
(E66)_{m_1 m_2 m_3}{}^{\; n_1 n_2 n_3} &= g^{1/2} \delta^{n_1 n_2 n_3}_{m_1 m_2 m_3}, \\
(E76)^{q_1 q_2}{}^{\; n_1 n_2 n_3} &= \frac{1}{2 \sqrt{3}} g V^{q_1 q_2 n_1 n_2 n_3}, \\
(E86)_{x}{}^{\; n_1 n_2 n_3} &= \frac{\sqrt{3}}{2 \sqrt{2}} g \left( U^{[n_1 n_2} \delta^{n_3]}_{x} + \frac{1}{6} X^{n_1 n_2 n_3}{}_{x} \right),
\end{align} 
\begin{align}
(E77)^{q_1 q_2}{}_{\; r_1 r_2} &=  g \delta^{q_1 q_2}_{r_1 r_2}, \\
(E87)_{x}{}_{\; r_1 r_2} &= \frac{1}{\sqrt{2}} g C_{x r_1 r_2}, \\
(E88)_{x}{}^{\; y} &=  g \delta^{y}_{x}.
\end{align}

\begin{sidewaystable}
\centering
\begin{tabular}{l}
$
\begin{pmatrix}
 (E11)^{a}{}_{\; b} & 0 & 0 & 0 & 0 & 0 & 0 & 0 \\
&&&&&&&\\
&&&&&&&\\
 (E21)_{d_1 d_2}{}_{\; b} & (E22)_{d_1 d_2}{}^{\; e_1 e_2} & 0 & 0 & 0 & 0 & 0 & 0 \\
&&&&&&&\\
&&&&&&&\\
 (E31)^{g_1 \dots g_3}{}_{\; b} & (E32)^{g_1 \dots g_3}{}^{\; e_1 e_2} & (E33)^{g_1 \dots g_3}{}_{\; h_1 \dots h_3} & 0 & 0 & 0 & 0 & 0 \\
&&&&&&&\\
&&&&&&&\\
 (E41)_{j_1}{}^{j_2}{}_{\; b} & (E42)_{j_1}{}^{j_2}{}^{\; e_1 e_2} & (E43)_{j_1}{}^{j_2}{}_{\; h_1 \dots h_3} & (E44)_{j_1}{}^{j_2}{}^{\; k_1}{}_{k_2} & 0 & 0 & 0 & 0 \\
&&&&&&&\\
&&&&&&&\\
 (E51)_{\; b} & (E52)^{\; e_1 e_2} & (E53)_{\; h_1 \dots h_3} & (E54)^{\;k_1}{}_{k_2} & E55 & 0 & 0 & 0 \\
&&&&&&&\\
&&&&&&&\\
 (E61)_{m_1 \dots m_3}{}_{\; b} & (E62)_{m_1 \dots m_3}{}^{\; e_1 e_2} & (E63)_{m_1 \dots m_3}{}_{\; h_1 \dots h_3} & (E64)_{m_1 \dots m_3}{}^{\; k_1}{}_{k_2} & (E65)_{m_1 \dots m_3} & (E66)_{m_1 \dots m_3}{}^{\; n_1 \dots n_3} & 0 & 0 \\
&&&&&&&\\
&&&&&&&\\
(E71)^{q_1 q_2}{}_{\; b} & (E72)^{q_1 q_2}{}^{\; e_1 e_2} & (E73)^{q_1 q_2}{}_{\; h_1 \dots h_3} & (E74)^{q_1 q_2}{}^{\; k_1}{}_{k_2} & (E75)^{q_1 q_2} & (E76)^{q_1 q_2}{}^{\; n_1 \dots n_3} & (E77)^{q_1 q_2}{}_{\; r_1 r_2} & 0 \\
&&&&&&&\\
&&&&&&&\\
 (E81)_{x}{}_{\; b} & (E82)_{x}{}^{\; e_1 e_2} & (E83)_{x}{}_{\; h_1 \dots h_3} & (E84)_{x}{}^{\; k_1}{}_{k_2} & (E85)_{x} & (E86)_{x}{}^{\; n_1 \dots n_3} & (E87)_{x}{}_{\; r_1 r_2} & (E88)_{x}{}^{\; y}
\end{pmatrix}
$ \vspace{6mm}
\end{tabular} 
\captionsetup{tablename=Figure}
\caption{The inverse generalised vielbein}
\label{tableinvviel} \end{sidewaystable} 

\newpage

\section{Calculation of potential}
\label{Vcal}

The potential of the canonical formulation of eleven-dimensional supergravity is given by
\begin{gather}
 V= \frac{1}{240} M^{MN} \partial_{M} M^{KL} \partial_{N} M_{KL} 
- \frac{1}{2} M^{MN} \partial_{N} M^{KL} \partial_{L} M_{MK}
- \frac{1}{496} M^{KL} \partial_{M} M^{MN} \partial_{N} M_{KL} \notag \\
+ \frac{23}{15(248)^2} M^{MN} (M^{KL} \partial_{M} M_{KL})( M^{RS} \partial_{N} M_{RS}),
\label{appV}
\end{gather}
where $M_{AB}$ is the generalised metric, \eqref{Mdef}, found from the non-realisation of the $E_{8}$ motion group and $M^{MN}$ is its inverse. The indices run from 1 to 248 and represent the adjoint representation of $E_{8}.$ In the decomposition of this representation by SL(8) irreducible representations,
$$\mathbf{248} = \mathbf{8} \oplus \mathbf{28} \oplus \mathbf{56} \oplus \mathbf{63} \oplus \mathbf{1} \oplus \mathbf{56} \oplus \mathbf{28} \oplus \mathbf{8},$$
we find the eight usual spatial directions along which the duality is acting along with 240 other directions that correspond to winding modes of branes. To produce a usual supergravity description from the duality-invariant description, from now on we take all the supergravity fields $g_{ab}, C_{abc}, C_{a_1 \dots a_6}, C_{a_1 \dots a_8, b}$ to be independent of the winding coordinates. Lowercase Latin indices are spatial coordinates and run from 1 to 8.

The coefficients in equation \eqref{appV} are fixed by requiring usual diffeomorphism invariance. Equivalently, they are fixed by requiring that when the gauge fields are zero the potential reduces to the Ricci scalar of metric $g.$ We now find what the potential is in terms of the supergravity fields. 

Since $M^{AB}$ is the matrix inverse of the generalised metric,
\begin{equation}
 M^{AB} = G^{CD} E^{A}{}_{C} E^{B}{}_{D},
\label{Minvdef}
\end{equation}
where $E^{A}{}_{B}$ is the inverse of the generalised vielbein, $$E^{A}{}_{B} L^{B}{}_{C}= \delta^{A}_{C}= L^{A}{}_{B} E^{B}{}_{C}$$ and
\begin{equation} 
G^{AB} = \textup{diag}(g^{ab}, g_{d_1 d_2, e_1 e_2}, g^{g_1 \dots g_3, h_1 \dots h_3}, g_{j_1 k_1} g^{j_2 k_2} - \frac{1}{8} \delta_{j_1}^{j_2} \delta_{k_1}^{k_2}, 
1, g_{m_1 \dots m_3, n_1 \dots n_3}, g^{q_1 q_2, r_1 r_2}, g_{x y}).
\label{Ginvdef}
\end{equation}
is the inverse of 
\begin{equation} 
G_{AB} = \textup{diag}(g_{ab}, g^{d_1 d_2, e_1 e_2}, g_{g_1 \dots g_3, h_1 \dots h_3}, g^{j_1 k_1} g_{j_2 k_2} - \frac{1}{8} \delta^{j_1}_{j_2} \delta^{k_1}_{k_2}, 
1, g^{m_1 \dots m_3, n_1 \dots n_3}, g_{q_1 q_2, r_1 r_2}, g^{x y}).
\label{Gdef2}
\end{equation} 
Using the equation \eqref{Mdef}, 
\begin{equation}
 M^{AB} = G^{CD} E^{A}{}_{C} E^{B}{}_{D},
\end{equation}
and equation \eqref{Minvdef} it is easy to show that
\begin{align}
M^{MN} \partial_{M} M^{KL} \partial_{N} M_{KL} &= 4 g^{ab} (\partial_{a} E^{K}{}_{C}) G^{CE} L^{F}{}_{K} (\partial_{b} G_{EF}) + 2 g^{ab} (\partial_{a} E^{K}{}_{C} ) (\partial_{b} L^{C}{}_{K})\notag \\
& \qquad  - 2 g^{ab}  G^{CD} G_{FG} L^{F}{}_{K} L^{G}{}_{L} (\partial_{a} E^{K}{}_{C} ) (\partial_{b} E^{L}{}_{D}) \notag \\
& \qquad \qquad+  g^{ab} (\partial_{a} G^{EF} ) (\partial_{b} G_{EF})
\label{term1}
\end{align}
and
\begin{align}
M^{MN} \partial_{N} M^{KL} \partial_{L} M_{MK} = - g^{ab} g^{cd} G_{KL}  L^{K}{}_{N} L^{L}{}_{M} (\partial_{b} E^{N}{}_{c} ) (\partial_{d} E^{M}{}_{a}) + g^{ab} (\partial_{b} g^{cd} ) (\partial_{d} g_{ac}).
\label{term2}
\end{align}
Furthermore, 
$$M^{CD} \partial_{a} M_{CD} = -248 \, g^{cd} \partial_{a} g_{cd},$$ 
hence
\begin{align}
M^{KL} \partial_{M} M^{MN} \partial_{N} M_{KL} = -248 (\partial_{a} g^{ab}) ( g^{cd} \partial_{b} g_{cd} ) 
\label{term3}
\end{align}
and 
\begin{align}
M^{MN} (M^{KL} \partial_{M} M_{KL})( M^{RS} \partial_{N} M_{RS}) = (248)^2 g^{ab} (g^{cd} \partial_{a} g_{cd}) (g^{ef} \partial_{b} g_{ef}).
\label{term4}
\end{align}

A simple calculation using the components of $G^{AB}$ and $G_{AB},$ equations \eqref{Ginvdef} and \eqref{Gdef2}, and the components of $L^{A}{}_{B}$ and $E^{A}{}_{B}$ given in appendix \ref{appviel} shows that 
\begin{gather}
g^{ab} (\partial_{a} E^{K}{}_{C}) G^{CE} L^{F}{}_{K} (\partial_{b} G_{EF}) = 6 g^{ab} (g^{cd} \partial_{a} g_{cd}) (g^{ef} \partial_{b} g_{ef}), \label{1term1} \\
g^{ab} (\partial_{a} E^{K}{}_{C} ) (\partial_{b} L^{C}{}_{K}) = -80 g^{ab} (g^{cd} \partial_{a} g_{cd}) (g^{ef} \partial_{b} g_{ef}), \label{1term2}\\
g^{ab} (\partial_{a} G^{EF} ) (\partial_{b} G_{EF}) = 60 g^{ab} (\partial_{a} g^{cd}) (\partial_{b} g_{cd}) - 12 g^{ab} (g^{cd} \partial_{a} g_{cd}) (g^{ef} \partial_{b} g_{ef}). \label{1term4}
\end{gather}
In particular, note that these are independent of the form fields and only depend on the metric $g.$ This is because $G_{AB}$ and $G^{AB}$ only depend on $g_{ab}.$ Moreover, $G_{AB}$ and $G^{AB}$ are diagonal and $L$ and $E$ are lower triangular so
$$(\partial_{a} E^{K}{}_{C}) G^{CF} L^{E}{}_{K} (\partial_{b} G_{EF}) \quad \textup{ and } \quad (\partial_{a} E^{K}{}_{C} ) (\partial_{b} L^{C}{}_{K})$$
only depend on the diagonal elements of $L$ and $E$ which are proportional to determinant of $g_{ab}.$

To calculate 
\begin{equation}
 g^{ab}  G^{CD} G_{FG} L^{F}{}_{K} L^{G}{}_{L} (\partial_{a} E^{K}{}_{C} ) (\partial_{b} E^{L}{}_{D})
\label{term1hard}
\end{equation}
in equation \eqref{term1} and 
\begin{equation}
 g^{ab} g^{cd} G_{KL}  L^{K}{}_{M} L^{L}{}_{N} (\partial_{b} E^{M}{}_{c} ) (\partial_{d} E^{N}{}_{a})
\label{term2hard}
\end{equation}
in equation \eqref{term2}, we note that the building block of both these terms is 
\begin{equation}
D_{u}{}^{A}{}_{B} = L^{A}{}_{C} (\partial_{u} E^{C}{}_{B}).
\label{D}
\end{equation}
The components of $D,$ see figure \ref{figD}, are given at the end of this appendix.

The evaluation of the components of $D$ requires use of identities such as 
\begin{gather}
 C_{a_1 a_2 b} V^{b c_1 \dots c_4} = 2 X^{[c_1 c_2 c_3}{}_{[a_1} \delta^{c_4]}_{a_2]},
\label{CViden}
\end{gather}
where $X$ is defined in equation \eqref{Xdef}. This identity is proved by using equation \eqref{Vdef} to write $C$ as a Hodge dual of $V$ in the expression above and $V$ in terms of $C.$ Then the two epsilon tensors are contracted to give a Kronecker delta. Finally using $$   C_{a_1 a_2 a_3} V^{c_1 c_2 a_1 a_2 a_3} = \frac{1}{3!} \epsilon^{c_1 c_2 a_1 a_2 a_3 b_1 b_2 b_3} C_{[a_1 a_2 a_3} C_{b_1 b_2 b_3]} =0,$$ we find the relation given above, equation \eqref{CViden}. Other useful identities are 
\begin{align}
  g^{-1/2} C_{a_1 a_2 a_3} \partial_u \left( g^{1/2} V^{c_1 c_2 a_1 a_2 a_3} \right)= - V^{c_1 c_2 b_1 b_2 b_3} \partial_u C_{b_1 b_2 b_3},
\label{VdCiden}
\end{align}
\begin{align}
  g^{-1/2} C_{a b c_1 \dots a_4} \partial_u \left( g^{1/2} U^{a b} \right)= U^{a b} \partial_u C_{a b c_1 \dots c_4},
\label{UdC6iden}
\end{align}
\begin{align}
  g^{-1/2} C_{a_1 a_2 b} \partial_u \left( g^{1/2} V^{c_1 c_2 c_3 a_1 a_2} \right)= V^{c_1 c_2 c_3 u_1 u_2} \partial_u C_{u_1 u_2 b} - V^{u_1 u_2 u_3 [c_1 c_2} \delta^{c_3]}_{b} \partial_u C_{u_1 u_2 u_3}.
\label{VdCiden2}
\end{align}
Note that $g^{1/2} \epsilon^{a_1 \dots a_8}$ is the alternating symbol
$$
\eta^{a_1 \dots a_8} = \begin{cases}
                        1 & \textup{ for } (a_1 \dots a_8) = \textup{ positive permutations of } (12345678) \\
 -1 & \textup{ for } (a_1 \dots a_8) = \textup{ negative permutations of } (12345678) \\
0 & \textup{ otherwise}
                       \end{cases},
$$
hence $$\partial_u \left( g^{1/2} \epsilon^{a_1 \dots a_8} \right) =0. $$ Furthermore, our convention for the contraction of two epsilon tensors is 
$$ \epsilon^{a_1 \dots a_i b_{i+1} \dots b_8} \epsilon_{c_1 \dots c_i b_{i+1} \dots b_8} = i! (8-i)! \delta^{a_1 \dots a_i}_{c_1 \dots c_i}.
$$

As an example, consider the evaluation of $D31,$
\begin{align*}
 (D31)_{u}{}^{g_1 g_2 g_3}{}_{b} &= (L31)^{g_1 g_2 g_3}{}_{c} \partial_{u}( E11)^{c}{}_{b} + (L32)^{g_1 g_2 g_3}{}^{f_1 f_2} \partial_{u} (E21)_{f_1 f_2}{}_{b} \qquad \qquad \qquad \\
&\qquad \qquad \qquad \qquad \qquad \qquad \qquad \qquad  + (L33)^{g_1 g_2 g_3}{}_{i_1 i_2 i_3} \partial_{u} (E31)^{i_1 i_2 i_3}{}_{b}.
\end{align*} 
Note that since $L$ is lower triangular there are only three terms contributing to $D31.$ The components of $L$ and $E$ can be read from appendix \ref{appviel} and inserted into the expression above \footnote{ This can either be done by hand or using the computer algebra software Cadabra \cite{cadabra}.}
\begin{align*}
 (D31)_{u}{}^{g_1 g_2 g_3}{}_{b} &= \frac{1}{2 \sqrt{6}} V^{g_1 g_2 g_3 f_1 f_2} \partial_{u} C_{f_1 f_2 b}  + \frac{\sqrt{3}}{2 \sqrt{2}} g^{-1/2} \partial_{u}   \left( g^{1/2} \delta^{[g_1}_{b} U^{g_2 g_3]} - \frac{1}{6} g^{1/2} X^{g_1 \dots g_3}{}_{b} \right)\\
&= \frac{1}{2 \sqrt{6}} V^{g_1 g_2 g_3 f_1 f_2} \partial_{u} C_{f_1 f_2 b}  + \frac{\sqrt{3}}{2 \sqrt{2}} g^{-1/2} \partial_{u}   \left( g^{1/2} \delta^{[g_1}_{b} U^{g_2 g_3]}\right) \\
& \qquad- \frac{1}{4 \sqrt{6}} g^{-1/2} C_{f_1 f_2 b} \partial_u \left( g^{1/2} V^{g_1 \dots g_3 f_1 f_2} \right) - \frac{1}{4 \sqrt{6}}   V^{g_1 \dots g_3 f_1 f_2}  \partial_u C_{f_1 f_2 b},
\end{align*} 
where we have used the definition of $X$ given in equation \eqref{Xdef}. Now, upon using identity \eqref{VdCiden2}, this reduces to 
$$(D31)_{u \;}{}^{g_1 g_2 g_3}{}_{\; b} = \frac{\sqrt{3}}{2 \sqrt{2}} g^{-1/2} \partial_{u} \left( g^{1/2} \delta^{[g_1}_{b} U^{g_2 g_3]} \right) + \frac{1}{4 \sqrt{6}} V^{t_1 t_2 t_3 [g_1 g_2} \delta^{g_3]}_{b} \partial_{u} C_{t_1 t_2 t_3}.$$
With the exception of $D61, D71, D81$ and $ D72,$ the other components can be simply derived using identities \eqref{CViden}--\eqref{VdCiden2}.

Showing that $D61, D71, D81$ and $ D72,$ vanish is not straightforward and involves the repeated use of identities \eqref{CViden}--\eqref{VdCiden2}. Expanding the $D61$ component we find thirteen terms of the form $$C^{(6)} V \partial C^{(3)}, \;\; C^{(6)} C^{(3)} \partial V, \;\;  U C^{(3)} \partial C^{(3)}, \;\; C^{(3)} C^{(3)} V \partial C^{(3)} \; \textup{ and } \; C^{(3)} C^{(3)} C^{(3)}  \partial V,$$ where $C^{(3)}$ and $C^{(6)}$ denote the 3 and 6-form, respectively. Expressing $C^{(6)} V \partial C^{(3)}$ and $C^{(6)} C^{(3)} \partial V$ as terms of the form $ U C^{(3)} \partial C^{(3)},$ it is easy to see that terms involving the 6-form cancel among each other. Further, writing terms of the form $C^{(3)} C^{(3)} V \partial C^{(3)}$ and $C^{(3)} C^{(3)} C^{(3)}  \partial V$ as the epsilon tensor multiplied by terms of the form $C^{(3)} V V \partial C^{(3)}$ we find that 
$$  (D61)_{u \;}{}_{m_1 m_2 m_3}{}_{\; b} =  - \frac{7}{16 (5!) \sqrt{6}} \epsilon_{c_1 \dots c_5 m_1 m_2 m_3} C_{a_1 a_2 b} V^{[c_1 \dots c_5} V^{d_1 d_2 d_3 a_1] a_2} \partial_{u} C_{d_1 d_2 d_3}.$$ This vanishes because an antisymmetrisation over nine indices in eight dimensions is zero. 
Similarly, $D71, D81$ and $ D72,$ also vanish upon repeated use of identities \eqref{CViden}--\eqref{VdCiden2}.

Given the components of $D$ it is now straightforward to evaluate expressions \eqref{term1hard} and \eqref{term2hard}. In terms of the supergravity fields these terms are 
\begin{align}
& g^{ab}  G^{CD} G_{FG} L^{F}{}_{K} L^{G}{}_{L} (\partial_{a} E^{K}{}_{C} ) (\partial_{b} E^{L}{}_{D})\\ = & g^{a b} G^{CD} G_{FG} D_{a}{}^{F}{}_{C} D_{b}{}^{G}{}_{D} \notag \\
= & 80 g^{ab} (g^{cd} \partial_{a} g_{cd}) (g^{ef} \partial_{b} g_{ef})  + 10 g^{ab} g^{c_{1} c_2 c_3, d_{1} d_{2} d_{3}} (\partial_{a} C_{c_{1} \dots c_{3}})(\partial_{b} C_{d_{1} \dots d_{3}}) \notag \\
& \quad + \frac{1}{48} g^{ab} g^{c_{1} \dots c_6, d_{1} \dots d_{6}} (\partial_{a} C_{c_{1} \dots c_{6}} - 20 C_{c_1 c_2 c_3} \partial_{a} C_{c_4 c_5 c_6})(\partial_{b} C_{d_{1} \dots d_{6}} - 20 C_{d_1 d_2 d_3} \partial_b C_{d_4 d_5 d_6})  \notag \\
&\qquad + \frac{15}{8!} g^{ab} g^{cd} g^{e_1 \dots e_8, f_1 \dots f_8} F_{a, e_1 \dots e_8, c} F_{b, f_1 \dots f_8, d}, 
\label{1term3}
\end{align}
where 
\begin{equation}
 F_{a, e_1 \dots e_8, b} = \partial_{a} C_{e_1 \dots e_8, b} - 28 C_{[e_1 \dots e_6|} \partial_{a} C_{|e_7 e_8] b} - \frac{560}{3} C_{b [ e_1 e_2} C_{e_3 e_4 e_5|} \partial_a C_{|e_6 e_7 e_8]}.
\label{f118def}
\end{equation}
Similarly,
\begin{align}
 & g^{ab} g^{cd} G_{KL}  L^{K}{}_{M} L^{L}{}_{N} (\partial_{b} E^{M}{}_{c} ) (\partial_{d} E^{N}{}_{a}) \notag \\ 
 =& g^{ab} g^{cd} G_{KL} D_b{}^{K}{}_{c} D_d{}^{L}{}_{a} \notag \\
=& \frac{1}{2} g^{a d_{1}} g^{c_{1} c_{2} c_3, b d_2 d_3} (\partial_{a} C_{c_{1} \dots c_{3}})(\partial_{b} C_{d_{1} \dots d_{3}}) 
- \frac{2}{8!} g^{a_1 \dots a_7, b_1 \dots b_7} F^{(7)}_{a_1 \dots a_7} F^{(7)}_{b_1 \dots b_7} \notag \\
& \;\;  + \frac{1}{4(6!)} g^{ab} g^{c_{1} \dots c_6, d_{1} \dots d_{6}} (\partial_{a} C_{c_{1} \dots c_{6}} - 20 C_{c_1 c_2 c_3} \partial_{a} C_{c_4 c_5 c_6})(\partial_{b} C_{d_{1} \dots d_{6}} - 20 C_{d_1 d_2 d_3} \partial_b C_{d_4 d_5 d_6}) \notag \\
& \quad + \frac{1}{4(8!)} g^{ab} g^{cd} g^{e_1 \dots e_8, f_1 \dots f_8} F_{a, e_1 \dots e_8, c} F_{b, f_1 \dots f_8, d} 
+ \frac{1}{4(8!)} g^{ad} g^{bc} g^{e_1 \dots e_8, f_1 \dots f_8} F_{a, e_1 \dots e_8, c} F_{b, f_1 \dots f_8, d},
\label{2term1} 
\end{align}
 where $F_{a, e_1 \dots e_8, b} $ is as in equation \eqref{f118def} and 
\begin{equation}
 F^{(7)}_{a_1 \dots a_7} = 7 \left( \partial_{[a_1} C_{a_2 \dots a_7]} + 20 C_{[a_1 a_2 a_3} \partial_{a_4} C_{a_5 a_6 a_7]} \right).
\label{f7def} 
\end{equation}

Therefore, using equations \eqref{term1}, \eqref{1term1}--\eqref{1term4} and \eqref{1term3}, 
\begin{align}
& M^{MN} \partial_{M} M^{KL} \partial_{N} M_{KL} \notag \\
= & 60 g^{ab} \partial_a g^{cd} \partial_b g_{cd} - 308 g^{ab} (g^{cd} \partial_{a} g_{cd}) (g^{ef} \partial_{b} g_{ef})  -20 g^{ab} g^{c_{1} c_2 c_3, d_{1} d_{2} d_{3}} (\partial_{a} C_{c_{1} \dots c_{3}})(\partial_{b} C_{d_{1} \dots d_{3}}) \notag \\
& \;\; - \frac{1}{24} g^{ab} g^{c_{1} \dots c_6, d_{1} \dots d_{6}} (\partial_{a} C_{c_{1} \dots c_{6}} - 20 C_{c_1 c_2 c_3} \partial_{a} C_{c_4 c_5 c_6})(\partial_{b} C_{d_{1} \dots d_{6}} - 20 C_{d_1 d_2 d_3} \partial_b C_{d_4 d_5 d_6})  \notag \\
&\qquad - \frac{30}{8!} g^{ab} g^{cd} g^{e_1 \dots e_8, f_1 \dots f_8} F_{a, e_1 \dots e_8, c} F_{b, f_1 \dots f_8, d},
\label{term1eval}
\end{align}
and from equations \eqref{term2} and \eqref{2term1}
\begin{align}
& M^{MN} \partial_{N} M^{KL} \partial_{L} M_{MK} \notag \\
&= g^{ab} (\partial_{b} g^{cd} ) (\partial_{d} g_{ac}) - \frac{1}{2} g^{a d_{1}} g^{c_{1} c_{2} c_3, b d_2 d_3} (\partial_{a} C_{c_{1} \dots c_{3}})(\partial_{b} C_{d_{1} \dots d_{3}}) 
+ \frac{2}{8!} g^{a_1 \dots a_7, b_1 \dots b_7} F^{(7)}_{a_1 \dots a_7} F^{(7)}_{b_1 \dots b_7} \notag \\
& \;\;  - \frac{1}{4(6!)} g^{ab} g^{c_{1} \dots c_6, d_{1} \dots d_{6}} (\partial_{a} C_{c_{1} \dots c_{6}} - 20 C_{c_1 c_2 c_3} \partial_{a} C_{c_4 c_5 c_6})(\partial_{b} C_{d_{1} \dots d_{6}} - 20 C_{d_1 d_2 d_3} \partial_b C_{d_4 d_5 d_6}) \notag \\
& \quad - \frac{1}{4(8!)} g^{ab} g^{cd} g^{e_1 \dots e_8, f_1 \dots f_8} F_{a, e_1 \dots e_8, c} F_{b, f_1 \dots f_8, d} 
- \frac{1}{4(8!)} g^{ad} g^{bc} g^{e_1 \dots e_8, f_1 \dots f_8} F_{a, e_1 \dots e_8, c} F_{b, f_1 \dots f_8, d}.
\label{term2eval}
\end{align}

Finally, putting together equations \eqref{term3}, \eqref{term4}, \eqref{term1eval}, \eqref{term2eval}, in terms of the supergravity fields the potential, \eqref{appV}, is
\begin{align}
 V &= \frac{1}{4} g^{ab} \partial_{a} g^{cd} \partial_b g_{cd} - \frac{1}{2} g^{ab} \partial_{b} g^{cd} \partial_d g_{ac} + \frac{1}{2} (\partial_{a} g^{ab}) ( g^{cd} \partial_{b} g_{cd} ) + \frac{1}{4} g^{ab} (g^{cd} \partial_{a} g_{cd}) (g^{ef} \partial_{b} g_{ef}) \notag \\
& \qquad -\frac{1}{12} g^{ab} g^{c_1 c_2 c_3, d_1 d_2 d_3} \partial_a C_{c_1 c_2 c_3} \left(  \partial_{b} C_{d_1 d_2 d_3} - 3 \partial_{d_1} C_{b d_2 d_3} \right) - \frac{1}{8!} g^{a_1 \dots a_7, b_1 \dots b_7} F^{(7)}_{a_1 \dots a_7} F^{(7)}_{b_1 \dots b_7} \notag \\
& \qquad \;\;\;\quad  + \frac{1}{8(8!)} g^{ad} g^{bc} g^{e_1 \dots e_8, f_1 \dots f_8} F_{a, e_1 \dots e_8, c} F_{b, f_1 \dots f_8, d}.
\end{align}
The first term is the Ricci scalar of metric $g,$ up to integration by parts. This expected because the coefficients of the terms in $V,$ equation \eqref{appV}, were fixed so that the Ricci scalar would be recovered when all other fields are zero. However, the potential also gives the dynamics of the other fields as well. Defining
$$F^{(4)}_{a_1 \dots a_4} = 4 \partial_{[a_1} C_{a_2 a_3 a_4]},$$
\begin{align}
 V &= R(g) -\frac{1}{48} g^{a_1 \dots a_4, b_1 \dots b_4} F^{(4)}_{a_1 \dots a_4} F^{(4)}_{b_1 \dots b_4} - \frac{1}{8!} g^{a_1 \dots a_7, b_1 \dots b_7} F^{(7)}_{a_1 \dots a_7} F^{(7)}_{b_1 \dots b_7} \notag \\
& \qquad \qquad\qquad \qquad\qquad \qquad \qquad\qquad +  \frac{1}{8(8!)} g^{ad} g^{bc} g^{e_1 \dots e_8, f_1 \dots f_8} F_{a, e_1 \dots e_8, c} F_{b, f_1 \dots f_8, d}.
\end{align}

\subsection{Components of $D_{u \;}{}^{A}{}_{\; B}$}

The components of $$D_{u \;}{}^{A}{}_{\; B} = L^{A}{}_{C} \partial_{u} E^{C}{}_{B}$$ are given below:
\begin{align}
 (D11)_{u \;}{}^{a}{}_{\; b} &= 0, \\
 (D21)_{u \;}{}_{d_1 d_2}{}_{\; b} &= \frac{1}{\sqrt{2}} \partial_{u} C_{d_1 d_2 b}, \\
 (D31)_{u \;}{}^{g_1 g_2 g_3}{}_{\; b}& = \frac{\sqrt{3}}{2 \sqrt{2}} g^{-1/2} \partial_{u} \left( g^{1/2} \delta^{[g_1}_{b} U^{g_2 g_3]} \right) + \frac{1}{4 \sqrt{6}} V^{t_1 t_2 t_3 [g_1 g_2} \delta^{g_3]}_{b} \partial_{u} C_{t_1 t_2 t_3}, \\
 (D41)_{u \;}{}_{j_1}{}^{j_2}{}_{\; b} = & - \frac{1}{2} \delta^{j_2}_{b} \left( g^{-1/2} \partial_{u} (g^{1/2} Y_{j_1}) -\frac{1}{2} U^{t_1 t_2} \partial_{u} C_{j_1 t_1 t_2} + \frac{1}{36} X^{t_1 t_2 t_3}{}_{j_1} \partial_{u} C_{t_1 t_2 t_3} \right) \notag \\
  & \; \; + \frac{1}{16} \delta^{j_2}_{j_1} \left( g^{-1/2} \partial_{u} (g^{1/2} Y_{b}) -\frac{1}{2} U^{t_1 t_2} \partial_{u} C_{b t_1 t_2} + \frac{1}{36} X^{t_1 t_2 t_3}{}_{b} \partial_{u} C_{t_1 t_2 t_3} \right), \\
 (D51)_{u \;}{}_{\; b} &= - \frac{3}{4 \sqrt{2}} \left( g^{-1/2} \partial_{u} (g^{1/2} Y_{b}) -\frac{1}{2} U^{t_1 t_2} \partial_{u} C_{b t_1 t_2} + \frac{1}{36} X^{t_1 t_2 t_3}{}_{b} \partial_{u} C_{t_1 t_2 t_3} \right), \\
 (D61)_{u \;}{}_{m_1 m_2 m_3}{}_{\; b} &=  0, \\
 (D71)_{u \;}{}^{q_1 q_2}{}_{\; b} &= 0, \\
(D81)_{u \;}{}_{x}{}_{\; b} &= 0,
\end{align}
\begin{align}
(D22)_{u \;}{}_{d_1 d_2}{}^{\; e_1 e_2} &= 0, \\
(D32)_{u \;}{}^{g_1 g_2 g_3}{}^{\; e_1 e_2} &= - \frac{1}{2 \sqrt{3}} g^{-1/2} \partial_{u} \left( g^{1/2} V^{g_1 g_2 g_3 e_1 e_2} \right), \\
(D42)_{u \;}{}_{j_1}{}^{j_2}{}^{\; e_1 e_2} =& \frac{1}{\sqrt{2}} \delta^{[e_2}_{j_1} \left( g^{-1/2} \partial_{u} (g^{1/2} U^{e_1] j_2}) + \frac{1}{6}  V^{ e_1] j_2 t_1 t_2 t_3 }  \partial_{u} C_{t_1 t_2 t_3} \right) \notag \\
  & \; \; - \frac{1}{8 \sqrt{2}} \delta^{j_2}_{j_1} \left( g^{-1/2} \partial_{u} (g^{1/2} U^{e_1 e_2}) -\frac{1}{6} V^{t_1 t_2 t_3 e_1 e_2}  \partial_{u} C_{t_1 t_2 t_3} \right), \\
(D52)_{u \;}{}^{\; e_1 e_2} &= - \frac{1}{24}  V^{e_1 e_2 t_1 t_2 t_3} \partial_{u} C_{t_1 t_2 t_3} - \frac{1}{4} g^{-1/2} \partial_u \left( g^{1/2} U^{e_1 e_2} \right), \\ 
(D62)_{u \;}{}_{m_1 m_2 m_3}{}^{\; e_1 e_2} = - \frac{\sqrt{3}}{2} \delta^{e_1 e_2}_{[m_1 m_2|} & \left( g^{-1/2} \partial_{u} (g^{1/2} Y_{|m_3]}) -\frac{1}{2} U^{t_1 t_2} \partial_{u} C_{|m_3] t_1 t_2} \right. \notag \\
& \left. \qquad \qquad\quad\qquad \qquad \qquad  + \frac{1}{36} X^{t_1 t_2 t_3}{}_{|m_3]} \partial_{u} C_{t_1 t_2 t_3} \right), \\
(D72)_{u \;}{}^{q_1 q_2}{}^{\; e_1 e_2} &= 0,
\end{align} 
\begin{align}
(D33)_{u \;}{}^{g_1 g_2 g_3}{}_{\; h_1 h_2 h_3} &= \frac{1}{2}  \delta^{g_1 g_2 g_3}_{h_1 h_2 h_3} (g^{-1} \partial_{u} g),  \\
(D43)_{u \;}{}_{j_1}{}^{\; j_2}{}_{\; h_1 h_2 h_3} &= \sqrt{\frac{3}{2}} \left(\partial_{u} C_{j_1 [h_1 h_2 } \delta^{j_2}_{h_{3}]} - \frac{1}{8} \delta^{j_2}_{j_1} \partial_{u} C_{h_1 h_2 h_3}  \right), \\
(D53)_{u \;}{}_{\; h_1 h_2 h_3} &=  \frac{1}{4 \sqrt{3}} \partial_{u} C_{h_1 h_2 h_3}, \\ 
(D63)_{u \;}{}_{m_1 m_2 m_3}{}_{\; h_1 h_2 h_3} &= - \frac{1}{12}  \partial_u  C_{m_1 m_2 m_3 h_1 h_2 h_3} + \frac{5}{3} C_{[m_1 m_2 m_3|} \partial_u C_{|h_1 h_2 h_3]},
\end{align} 
\begin{align}
(D44)_{u \;}{}_{j_1}{}^{j_2}{}^{\; k_1}{}_{k_2} &= \frac{1}{2} \left( \delta^{j_2}_{k_2} \delta^{k_1}_{j_1} - \frac{1}{8} \delta^{j_2}_{j_1} \delta^{k_1}_{k_2} \right) (g^{-1} \partial_{u} g), \\
(D54)_{u \;}{}^{\; k_1}{}_{k_2} &= 0,  \\ 
\end{align} 
\begin{align}
D55 &= \frac{1}{2} (g^{-1} \partial_{u} g),  \\ 
(D66)_{u \;}{}_{m_1 m_2 m_3}{}^{\; n_1 n_2 n_3} &= \frac{1}{2} \delta^{n_1 n_2 n_3}_{m_1 m_2 m_3} (g^{-1} \partial_{u} g), \\
(D77)_{u \;}{}^{q_1 q_2}{}_{\; r_1 r_2} &= \delta^{q_1 q_2}_{r_1 r_2} (g^{-1} \partial_{u} g), \\
(D88)_{u \;}{}_{x}{}^{\; y} &= \delta^{y}_{x} (g^{-1} \partial_{u} g).
\end{align}

\begin{sidewaystable}
\centering
\begin{tabular}{l}
{ \tiny  $
\begin{pmatrix}
 (D11)_{u \;}{}^{a}{}_{\; b} & 0 & 0 & 0 & 0 & 0 & 0 & 0 \\
&&&&&&&\\
&&&&&&&\\
 (D21)_{u \;}{}_{d_1 d_2}{}_{\; b} & (D22)_{u \;}{}_{d_1 d_2}{}^{\; e_1 e_2} & 0 & 0 & 0 & 0 & 0 & 0 \\
&&&&&&&\\
&&&&&&&\\
 (D31)_{u \;}{}^{g_1 \dots g_3}{}_{\; b} & (D32)_{u \;}{}^{g_1 \dots g_3}{}^{\; e_1 e_2} & (D33)_{u \;}{}^{g_1 \dots g_3}{}_{\; h_1 \dots h_3} & 0 & 0 & 0 & 0 & 0 \\
&&&&&&&\\
&&&&&&&\\
 (D41)_{u \;}{}_{j_1}{}^{j_2}{}_{\; b} & (D42)_{u \;}{}_{j_1}{}^{j_2}{}^{\; e_1 e_2} & (D43)_{u \;}{}_{j_1}{}^{j_2}{}_{\; h_1 \dots h_3} & (D44)_{u \;}{}_{j_1}{}^{j_2}{}^{\; k_1}{}_{k_2} & 0 & 0 & 0 & 0 \\
&&&&&&&\\
&&&&&&&\\
 (D51)_{u \;}{}_{\; b} & (D52)_{u \;}{}^{\; e_1 e_2} & (D53)_{u \;}{}_{\; h_1 \dots h_3} & (D54)_{u \;}{}^{\;k_1}{}_{k_2} & D55 & 0 & 0 & 0 \\
&&&&&&&\\
&&&&&&&\\
 (D61)_{u \;}{}_{m_1 \dots m_3}{}_{\; b} & (D62)_{u \;}{}_{m_1 \dots m_3}{}^{\; e_1 e_2} & (D63)_{u \;}{}_{m_1 \dots m_3}{}_{\; h_1 \dots h_3} & (D43)_{u \;}{}_{k_2}{}^{k_1}{}_{\; m_1 \dots m_3} & (D53)_{u \;}{}_{\; m_1 \dots m_3} & (D66)_{u \;}{}_{m_1 \dots m_3}{}^{\; n_1 \dots n_3} & 0 & 0 \\
&&&&&&&\\
&&&&&&&\\
(D71)_{u \;}{}^{q_1 q_2}{}_{\; b} & (D72)_{u \;}{}^{q_1 q_2}{}^{\; e_1 e_2} & -(D62)_{u \;}{}_{h_1 \dots h_3}{}^{\; q_1 q_2} & (D42)_{u \;}{}_{k_2}{}^{k_1}{}^{\; q_1 q_2} & (D52)_{u \;}{}^{\; q_1 q_2} & - (D32)_{u \;}{}^{n_1 \dots n_3}{}^{\; q_1 q_2} & (D77)_{u \;}{}^{q_1 q_2}{}_{\; r_1 r_2} & 0 \\
&&&&&&&\\
&&&&&&&\\
 (D81)_{u \;}{}_{x}{}_{\; b} & (D71)_{u \;}{}^{\; e_1 e_2}{}_{\; x} & (D61)_{u \;}{}_{h_1 \dots h_3}{}_{\; x} & -(D41)_{u \;}{}_{k2}{}^{k_1}{}_{\; x} & - (D51)_{u \;}{}_{\; x} & (D31)_{u \;}{}^{n_1 \dots n_3}{}_{\; x} & (D21)_{u \;}{}_{r_1r_2}{}_{\; x} & (D88)_{u \;}{}_{x}{}^{\; y}
\end{pmatrix}
$} \vspace{6mm}
\end{tabular} 
\captionsetup{tablename=Figure}
\caption{Components of $D_{u}{}^{A}{}_{B} = L^{A}{}_{C} (\partial_{u} E^{C}{}_{B})$}
\label{figD} \end{sidewaystable} 

\pagebreak

\bibliography{e8}
\bibliographystyle{utphys}
\end{document}